\documentclass[iop,usenatbib]{emulateapj}

\usepackage[english]{babel}
\usepackage{graphicx}
\usepackage{fleqn}
\usepackage{amssymb}
\usepackage{amsmath}
\usepackage{booktabs}
\usepackage{threeparttable}
\usepackage{enumerate}
\usepackage[breaklinks,colorlinks,
   urlcolor=blue,citecolor=blue,linkcolor=blue]{hyperref}
\usepackage{comment}

\usepackage{txfonts} 

\renewcommand{\S}{Section}
\newcommand{\F}{Fig.}

\newcommand{\ve}[1]{\mathbf{#1}}
\newcommand{\unit}[1]{\hat{\mathbf{#1}}}

\newcommand{\Ec}{E}

\newcommand{\EA}{\mathcal{E}} 
 
\newcommand{\TA}{\theta}

\newcommand{\msun}{\mathrm{M}_\odot}
\newcommand{\rsun}{\mathrm{R}_\odot}

\newcommand{\au}{\,\textsc{au}}

\renewcommand{\d}{\mathrm{d}}
\renewcommand{\a}{\mathrm{a}}
\newcommand{\orb}{\mathrm{orb}}

\newcommand{\Md}{M_\mathrm{d}}

\newcommand{\rAd}{\ve{r}_{\mathrm{A}_\mathrm{d}}}
\newcommand{\Ad}{{\mathrm{A}_\mathrm{d}}}
\newcommand{\Aa}{{\mathrm{A}_\mathrm{a}}}

\newcommand{\XL}{X_{\mathrm{L}}}
\newcommand{\XLz}{X_{\mathrm{L,0}}}

\newcommand{\rAa}{\ve{r}_{\mathrm{A}_\mathrm{a}}}
\newcommand{\rA}{\ve{r}_{\mathrm{A}}}
\usepackage{bm}

\begin{document}

\title{An analytic model for mass transfer in binaries with arbitrary eccentricity, with applications to triple-star systems}
\author{Adrian S. Hamers$^{1}$ and Fani Dosopoulou$^{2,3}$}
\affil{$^{1}$Institute for Advanced Study, School of Natural Sciences, Einstein Drive, Princeton, NJ 08540, USA \\
$^{2}$Princeton Center for Theoretical Science, Princeton University, Princeton, NJ 08544, USA \\
$^{3}$Department of Astrophysical Sciences, Princeton University, Princeton, NJ 08544, USA}
\email{hamers@ias.edu}

\begin{abstract}  
Most studies of mass transfer in binary systems assume circular orbits at the onset of Roche lobe overflow. However, there are theoretical and observational indications that mass transfer could occur in eccentric orbits. In particular, eccentricity could be produced via sudden mass loss and velocity kicks during supernova explosions, or Lidov-Kozai (LK) oscillations in hierarchical triple systems, or, more generally, secular evolution in multiple-star systems. However, current analytic models of eccentric mass transfer are faced with the problem that they are only well defined in the limit of very high eccentricities, and break down for less eccentric and circular orbits. This provides a major obstacle to implementing such models in binary and higher-order population synthesis codes, which are useful tools for studying the long-term evolution of a large number of systems. Here, we present a new analytic model to describe the secular orbital evolution of binaries undergoing conservative mass transfer. The main improvement of our model is that the mass transfer rate is a smoothly varying function of orbital phase, rather than a delta function centered at periapsis. Consequently, our model is in principle valid for any eccentricity, thereby overcoming the main limitation of previous works. We implement our model in an easy-to-use and publicly available code that can be used as a basis for implementations of our model into population synthesis codes. We investigate the implications of our model in a number of applications with circular and eccentric binaries, and triples undergoing LK oscillations.
\end{abstract}


\section{Introduction}
\label{sect:introduction}
A common evolutionary process in binary and multiple-star systems is the transfer of mass between stars. Mass transfer is thought to be responsible for a wide range of phenomena such as X-ray emission in low- and high-mass X-ray binaries (see, e.g., \citealt{1993ARA&A..31...93V,2006ARA&A..44...49R} for reviews), spin-up of neutron stars (e.g., \citealt{2008LRR....11....8L}), and cataclysmic variables (e.g., \citealt{2011ASPC..447....3K}). 

Since tides are generally thought to be efficient in close binary systems \citep{1977A&A....57..383Z,1981ARA&A..19..277S}, most theoretical studies of mass transfer assume that the orbit has circularized by the time of onset of mass transfer \citep{2002MNRAS.329..897H,2003ASPC..303..290P}. However, various processes are known to be able to excite significant eccentricity, even when tides are taken into account during the evolution of the binary. These processes include sudden mass loss and an imparted velocity kick during supernova explosions (e.g., \citealt{1983ApJ...267..322H,1995MNRAS.274..461B,1996ApJ...471..352K}), enhanced mass loss at periapsis \citep{2000A&A...357..557S,2008A&A...480..797B}, interactions with a massive circumbinary disk \citep{2013A&A...551A..50D,2014ApJ...797L..24A,2016ApJ...830....8R,2018arXiv181004676M}, or Lidov-Kozai (LK) oscillations (\citealt{1962P&SS....9..719L,1962AJ.....67..591K}; see \citealt{2016ARA&A..54..441N} for a review) if the binary is orbited by a third star (or, more generally, secular evolution in higher-order systems including, but not limited to, quadruple systems, e.g., \citealt{2013MNRAS.435..943P,2015MNRAS.449.4221H,2016MNRAS.459.2827H,2017MNRAS.470.1657H,2018MNRAS.474.3547G}). In particular, population synthesis studies of triple stars find that of the order of 10\% of systems undergo mass transfer in eccentric orbits at some point in their evolution \citep{2016ComAC...3....6T,2018A&A...610A..22T}. Similarly, mass transfer in eccentric orbits can be triggered by secular evolution in quadruple-star systems \citep{2018MNRAS.478..620H,2019MNRAS.482.2262H}. From an observational side, semidetached binaries are known to have nonzero eccentricities \citep{1999AJ....117..587P,2008A&A...480..797B,2013A&A...559A..54V,2014A&A...564A...1B}, as well as high-mass X-ray binaries \citep{2005A&AT...24..151R}, and post-AGB binaries \citep{1995A&A...293L..25V}.

Despite the relevance of nonzero eccentricity in mass transfer processes, binary population synthesis codes such as \textsc{StarTrack} \citep{2008ApJS..174..223B}, \textsc{BSE} \citep{2002MNRAS.329..897H} and the updated \textsc{binary\_c} \citep{2004MNRAS.350..407I,2006A&A...460..565I,2009A&A...508.1359I,2014A&A...563A..83C}, and \textsc{SeBa} \citep{1996A&A...309..179P,2012A&A...546A..70T}, enforce circular orbits at the onset of mass transfer. Nevertheless, the problem of mass loss/mass transfer has been studied for over half a century (e.g., \citealt{1956AJ.....61...49H,1963Icar....2..440H,1964AcA....14..241K,1964AcA....14..251P,1983ApJ...266..776M,1984ApJ...282..522M}), and has received more recent attention in numerical studies (e.g., \citealt{2005MNRAS.358..544R,2009MNRAS.395.1127C,2010ApJ...724..546S,2011ApJ...726...66L,2011ApJ...726...67L,2016MNRAS.455..462V,2017MNRAS.467.3556B}), as well as in (semi)analytical work \citep{2007ApJ...667.1170S,2009ApJ...702.1387S,2010ApJ...724..546S,2011MNRAS.417.2104V,2012MNRAS.422.1648V,2013MNRAS.435.2416V,2014MNRAS.437.1127V,2016ApJ...825...70D,2016ApJ...825...71D}. 

\citet{2007ApJ...667.1170S} and \citet{2016ApJ...825...71D} in particular derived equations for the secular (i.e., orbit-averaged) changes of the orbital elements due to mass transfer in eccentric binaries. For the case of Roche Lobe overflow (RLOF), they assumed that the mass transfer rate is a delta function centered at periapsis, i.e., the donor star transfers its mass in a burst at its closest approach to its companion. This assumption is physically reasonable in the limit of very high eccentricity. However, it is clearly no longer reasonable for less eccentric or even circular orbits. In the latter case, the mass transfer rate is expected to be constant during the orbit. Moreover, periapsis is no longer defined for a circular orbit. 

In the delta function model of eccentric RLOF of \citet{2007ApJ...667.1170S} and \citet{2016ApJ...825...71D}, the equations of motion state that the eccentricity time derivative is negative at zero eccentricity, provided that the mass ratio $q$ of the donor to the accretor mass is $q>1$ (and assuming point masses). This has the practical implication that, when solving the equations of motion (a set of first-order ordinary differential equations [ODEs]) numerically, the eccentricity becomes negative as the system evolves toward circularization (see \F\,\ref{fig:binary_circular} in \S\,\ref{sect:appl_bin:circ} below for an example). Clearly, this is an undesirable property, especially when the equations are to be implemented in population synthesis codes. The case $q<1$ is problematic as well, since for $q<1$ the eccentricity time derivative is positive at zero eccentricity, and the equations of motion predict a growing eccentricity, which ultimately leads to significant deviation from the expected evolution of the semimajor axis for strictly circular orbits (see equation~\ref{eq:a_dot_can} below and \F\,\ref{fig:binary_circular_low_q} for an example). 

In this paper, we present an analytic model for mass transfer in eccentric orbits that shares some of the same basic assumptions as those of \citet{2007ApJ...667.1170S} and \citet{2016ApJ...825...71D}, but assumes a physically motivated and more realistic model for the mass transfer rate. Consequently, our model is in principle valid for any eccentricity, including zero, eliminating the issues described above. We give explicit expressions for the orbit-averaged equations of motion and apply them to mass-transferring isolated binaries and triple systems. We also make an easy-to-use \textsc{Python} code publicly available (see \S\,\ref{sect:model:num} for the url) to quickly solve the equations of motion numerically, and which can be used as a basis for implementations of our model into binary (and higher-order multiplicity) population synthesis codes. 

The plan of the paper is as follows. The analytic model is presented in \S\,\ref{sect:model}. We give applications of the model in isolated binaries in \S\,\ref{sect:appl_bin}, and in triple systems undergoing LK oscillations in \S\,\ref{sect:appl_triple}. We discuss limitations of our model in \S\,\ref{sect:discussion}, and conclude in \S\,\ref{sect:conclusions}.

\section{The analytic model}
\label{sect:model}

Consider a binary with a donor star with mass $M_\d$ and radius $R$, and an accreting star with mass $M_\a$. Let the total mass be denoted with $M\equiv M_\d+M_\a$, the mass ratio with $q\equiv M_\d/M_\a$, the reduced mass with $\mu \equiv M_\d M_\a/M$, and the relative separation vector between the centers of mass of the stars with $\ve{r}\equiv \ve{R}_\a-\ve{R}_\d$, where $\ve{R}_\d$ and $\ve{R}_\a$ are the absolute position vectors of the centers of mass of the donor and accretor star, respectively (here, `absolute' means relative to an inertial reference frame). The orbital angular frequency vector is $\bm{\omega}_\orb$, which is directed along the orbital angular momentum vector, and which has a magnitude $\omega_\orb = \dot{\TA} = n \sqrt{1-e^2} \,(a/r)^2$, where $\TA$ is the true anomaly, the dot denotes the derivative with respect to time $t$, $n = \sqrt{GM/a^3}$ is the mean motion, and $a$ and $e$ are the orbital semimajor axis and eccentricity, respectively. 

Assume that the donor loses mass with a mass-loss rate $\dot{M}_\d$, and an absolute velocity $\ve{W}_\d$ at the position $\ve{r}_\Ad$ relative to the center of mass of the donor (i.e., at the Lagrangian point $L_1$ in the case of RLOF). The accretor accretes mass with a rate $\dot{M}_\a$, and absolute velocity $\ve{W}_\a$ at the position $\rAa$ relative to the center of mass of the accretor ($||\rAa||$ could be the accretor's radius in the case of direct impact accretion, or, e.g., the size of the accretion disk if it is present). The ejection/accretion velocities relative to the donor and accretor are $\ve{w}_\d \equiv \ve{W}_\d - \ve{V}_\d$ and $\ve{w}_\a \equiv \ve{W}_\a - \ve{V}_\a$, respectively, where $\ve{V}_\d\equiv \mathrm{d}\ve{R}_\d/\mathrm{d} t$ and $\ve{V}_\a\equiv \mathrm{d} \ve{R}_\a/\mathrm{d} t$ are the absolute velocities of the centers of mass of the donor and accretor, respectively.

\subsection{Perturbing acceleration due to mass transfer}
\label{sect:model:per}
\subsubsection{General equations of motion}
\label{sect:model:per:eom}

We adopt the approach in which the effects of mass transfer are treated as perturbations to the instantaneous (osculating) Kepler orbit of the binary. Note that another approach is to consider changes in the total binary orbital energy and angular momentum (e.g., \citealt{1956AJ.....61...49H,2008A&A...480..797B}). As derived by \citet{1969Ap&SS...3...31H} and re-derived by \citet{2007ApJ...667.1170S}, the acceleration of the relative position vector $\ve{r}$ can be written as
\begin{subequations}
\label{eq:EOM}
\begin{align}
\label{eq:EOM_l1} & \frac{\mathrm{d}^2 \ve{r}}{\mathrm{d} t^2} = -\frac{GM}{r^3} \ve{r} \\
\label{eq:EOM_l2} &\quad + \frac{\ve{f}_\a}{M_\a} - \frac{\ve{f}_\d}{M_\d}  \\
\label{eq:EOM_l3} &\quad + \frac{\dot{M}_\a}{M_\a} \left (\ve{w}_\a + \bm{\omega}_\orb \times \ve{r}_\Aa \right ) -  \frac{\dot{M}_\d}{M_\d} \left (\ve{w}_\d + \bm{\omega}_\orb \times \ve{r}_\Ad \right ) \\
\label{eq:EOM_l4}&\quad + \frac{\ddot{M}_\a}{M_\a} \ve{r}_\Aa - \frac{\ddot{M}_\d}{M_\d} \ve{r}_\Ad.
\end{align}
\end{subequations}
The first term after the equality in the first line (\ref{eq:EOM_l1}) is the Keplerian acceleration, and all other terms represent perturbations associated with mass transfer. The terms in the second line (\ref{eq:EOM_l2}) represent perturbations from the ejected mass on the orbit. These terms are generally hard to calculate analytically, and typically require numerical integration of the trajectories of the particles in the mass transfer stream (e.g., \citealt{1969Ap&SS...3..330H,2010ApJ...724..546S,2014A&A...570A..25D}). The terms in the third line (\ref{eq:EOM_l3}) are associated with the change of linear momentum of the accretor and donor due to mass transfer. The terms in the fourth line (\ref{eq:EOM_l4}) are due to the acceleration of the centers of mass of the accretor and donor. 

We note that equation~(\ref{eq:EOM}) was also derived by \citet{1983ApJ...266..776M,1984ApJ...282..522M}, although, due to an error made in the equations for the absolute position vectors of the components in terms of the relative separation, the acceleration term (line \ref{eq:EOM_l4}) was missing, as was pointed out in section 3.3 of \citet{2007ApJ...667.1170S}.

\subsubsection{Simplified equations -- overview}
\label{sect:model:per:simple_overview}
The general equations of motion, equation~(\ref{eq:EOM}), are hard to model analytically without detailed (hydrodynamical) simulations (see, e.g., \citealt{1987MNRAS.229..383E,2005MNRAS.358..544R,2009MNRAS.395.1127C,2010ApJ...724..546S,2011ApJ...726...66L,2011ApJ...726...67L,2016MNRAS.455..462V,2017MNRAS.467.3556B} for numerical/hydrodynamical studies). We make a number of simplifying assumptions in order to arrive at equations that are analytically tractable. These assumptions are summarized below.

\begin{enumerate}
\item We assume that the effects of the mass stream on the orbit are negligible, i.e., we set $\ve{f}_\a = \ve{f}_\d = \ve{0}$. 
\item We assume conservative mass transfer, i.e., $\dot{M}_\d = -\dot{M}_\a$, such that $\dot{M}=0$. 
\item We assume that the donor ejects mass at a relative velocity of $\ve{w}_\d = \dot{\ve{r}}$, and that the accretor accretes mass at a relative velocity of $\ve{w}_\a = - \dot{\ve{r}}$. 
\item We assume that $\ve{r}_\Ad$ and $\ve{r}_\Aa$ corotate with the orbit, i.e., they are proportional to $\unit{r}$. In addition, we take $\ve{r}_\Aa$ to have a constant magnitude, whereas we make two limiting assumptions of the magnitude of $\ve{r}_\Ad$: proportional to $r(t)$, and constant as a function of orbital phase.
\end{enumerate}

The third assumption in particular may seem peculiar. However, we show below in \S\,\ref{sect:model:per:simple_zero} that these assumptions, in the case of point masses with zero-size ejection/accretion radii ($\ve{r}_\Ad=\ve{r}_\Aa=\ve{0}$), lead to the canonical relation for the change of the semimajor axis due to conservative mass transfer in circular orbits.

\subsubsection{Simplified equations -- point masses and circular orbits}
\label{sect:model:per:simple_zero}
Here, we show that the assumptions of \S\,\ref{sect:model:per:simple_overview}, combined with $\ve{r}_\Ad=\ve{r}_\Aa=\ve{0}$, yield the canonical relation for mass transfer in circular orbits, i.e.,
\begin{align}
\label{eq:a_dot_can}
\frac{\dot{a}}{a} = -2\frac{\dot{\Md}}{\Md} \left (1-\frac{M_\d}{M_\a} \right ).
\end{align}
This relation is usually derived using conservation of the orbital angular momentum, $L_\mathrm{orb} = \mu \sqrt{GMa(1-e^2)}$, i.e., by setting
\begin{align}
\frac{\dot{L}_\mathrm{orb}}{L_\mathrm{orb}} = \frac{\dot{M}_\d}{M_\d} +  \frac{\dot{M}_\a}{M_\a} - \frac{1}{2} \frac{\dot{M}}{M} + \frac{1}{2} \frac{\dot{a}}{a} - \frac{e \dot{e}}{1-e^2}
\end{align}
to zero, combined with the assumptions of conservative mass transfer ($\dot{M}=0$), and circular orbits ($e=0$). Equation~(\ref{eq:a_dot_can}) states that the orbit shrinks when the donor is more massive than the accretor; when the mass ratio has reversed, the orbit expands (note that $\dot{M}_\d<0$). 

With the assumptions outlined above in \S\,\ref{sect:model:per:simple_overview} and setting $\ve{r}_\Ad=\ve{r}_\Aa=\ve{0}$, equation~(\ref{eq:EOM}) reduces to
\begin{align}
\label{eq:EOM_simple_zero} 
\frac{\mathrm{d}^2 \ve{r}}{\mathrm{d} t^2} &= -\frac{GM}{r^3} \ve{r} - \frac{\dot{M}_\d}{M_\d} \left (1-\frac{M_\d}{M_\a} \right )\dot{\ve{r}}.
\end{align}

Consider the secular orbital evolution implied by equation~(\ref{eq:EOM_simple_zero}). If the orbits are circular, then it is reasonable to assume that the mass transfer rate $\dot{M}_\d$ is constant as well. In addition, if the orbital timescale is short compared to the mass transfer timescale, $P_\mathrm{orb}\ll M_\d/\dot{M}_\d$, then we can assume that $M_\d$ and $M_\a$ are approximately constant during the orbit (i.e., the adiabatic approximation). The resulting secular semimajor axis change (see \S\,\ref{sect:model:av:pre} below for details) is
\begin{align}
\label{eq:a_dot_can_der}
\left \langle \frac{\dot{a}}{a} \right \rangle &= \frac{2a}{GM}\frac{-\dot{M}_\d}{M_\d}\left(1-\frac{M_\d}{M_\a}\right ) \left \langle \dot{\ve{r}}^2 \right \rangle=-2\frac{\dot{\Md}}{\Md} \left (1-\frac{M_\d}{M_\a} \right ).
\end{align}
We thus arrive at the canonical relation, equation~(\ref{eq:a_dot_can}). 

We note that equation~(\ref{eq:EOM_simple_zero}) is also commonly used in other studies of mass transfer (e.g., \citealt[eq. C98]{2006epbm.book.....E}; \citealt{2018MNRAS.480.3195K}).

\subsubsection{Simplified equations -- extended bodies with nonzero ejection/accretion radii}
\label{sect:model:per:simple_nonzero}
With the assumptions described in \S\,\ref{sect:model:per:simple_overview} and not setting the ejection/accretion radii to zero, the equations of motion (equation~\ref{eq:EOM}) can be written as
\begin{align}
\nonumber \frac{\mathrm{d}^2 \ve{r}}{\mathrm{d} t^2} &= -\frac{GM}{r^3} \ve{r} - \frac{\dot{M}_\d}{M_\d} \left (1-q \right )\dot{\ve{r}} - \frac{\dot{M}_\d}{M_\d} \bm{\omega}_\orb \times \left (\ve{r}_\Ad + q \, \rAa \right ) \\
\nonumber &\quad - \frac{\ddot{M}_\d}{M_\d} \left (\ve{r}_\Ad + q \, \ve{r}_\Aa \right ) \\
\label{eq:EOM_simple}
&\equiv -\frac{GM}{r^3} \ve{r} - \frac{\dot{M}_\d}{M_\d} \left (1-q \right )\dot{\ve{r}} - \frac{\dot{M}_\d}{M_\d} \bm{\omega}_\orb \times \rA -  \frac{\ddot{M}_\d}{M_\d} \rA,
\end{align}
where we introduced, for convenience, the short-hand notation
\begin{align}
\label{eq:rA}
\rA \equiv \ve{r}_\Ad + q\, \ve{r}_\Aa.
\end{align}
When averaging over the orbit below in \S\,\ref{sect:model:av}, we follow the adiabatic approximation and set $M_\d$ and the mass ratio $q$ to be constant. We do not assume a constant $\rA$, as is described in more detail below in \S\,\ref{sect:model:rA}.

Equation~(\ref{eq:EOM_simple}) is consistent with the equations of motion assumed by \citet{2007ApJ...667.1170S} and \citet{2016ApJ...825...71D}. In further steps below, we deviate from these works. In particular, we make different assumptions of the ejection/accretion radii (\S\,\ref{sect:model:rA}), and the instantaneous mass transfer rate (\S\,\ref{sect:model:rate}).

\subsection{Assumptions of the ejection/accretion radii}
\label{sect:model:rA}
The ejection/accretion positions $\ve{r}_\Ad$ and $\ve{r}_\Aa$ describe the locations of the ejected/accreted mass relative to the donor/accretor star. Seen from an inertial frame, the vectors $\ve{r}_\Ad$ and $\rAa$ rotate. Here, we assume that $\ve{r}_\Ad$ is aligned with the separation vector $\ve{r}$ between the two stars, and points towards the accretor, i.e., $\unit{r}_\Ad = \unit{r}$. We assume that the mass is accreted by the accretor on the same axis, but in the opposite direction, i.e., $\unit{r}_\Aa = - \unit{r}$. 

In general, the magnitudes of both $\ve{r}_\Ad$ and $\ve{r}_\Aa$ could be functions of the orbital phase. In the case of RLOF, $\ve{r}_\Ad$ is the location of the first Lagrangian point $L_1$, interpreted to be a function of orbital phase, and insofar as this point can be defined for eccentric orbits. In the limit that the dynamical timescale of the donor is much shorter than the timescale associated with the orbital angular velocity and donor rotation (also known as the {\it first approximation}, \citealt{1963ApJ...138.1112L}), \citet{2007ApJ...660.1624S} showed that the stationary point $L_1$ between the two stars can be determined according to the equation
\begin{align}
\label{eq:XL}
\frac{q}{\XL^2} - \frac{1}{(1-\XL)^2} - \XL (1+q) \mathcal{A}(\hat{\Omega},e,\TA) + 1 = 0.
\end{align}
Here, $\XL \equiv X/r$ is the location of $L_1$ relative to the donor's center of mass normalized to the orbital separation, $\hat{\Omega}\equiv \Omega/\omega_{\orb,\,\mathrm{P}}$ is the donor's spin frequency $\Omega$ normalized to $\omega_{\orb,\,\mathrm{P}} = n (1+e)^{1/2}/(1-e)^{3/2}$, the orbital angular frequency at periapsis, and 
\begin{align}
\label{eq:Adef}
\mathcal{A}(\hat{\Omega},e,\TA) = \frac{\hat{\Omega}^2 (1+e)^4}{(1+e \cos\TA)^3} = \hat{\Omega}^2 \frac{1+e}{(1-e)^3} \left ( \frac{r}{a} \right )^3.
\end{align}
Unfortunately, no (simple) analytic solutions exist for $\XL$ in equation~(\ref{eq:XL}) as a function of $q$ and $\mathcal{A}$. Instead, we make two limiting assumptions.
\begin{enumerate}
\item Negligible donor spin: $\hat{\Omega}\approx 0$.
\item Large mass ratio: $q\gg1$.
\end{enumerate}
In the first case, $\mathcal{A}\approx 0$, and we can neglect the associated term in equation~(\ref{eq:XL}). Consequently, $\XL$ is a function only of $q$. An analytic (although not very illuminating) solution for $\XL=\XLz(q)$ exists in this case, and is given explicitly in Appendix~\,\ref{app:XL}. Note that, although $\XL$ is not a function of orbital phase, the location of $L_1$ itself, $X = \XLz(q) r(t)$, varies along the orbit. 

In the second case, we can ignore any terms not involving $q$ in equation~(\ref{eq:XL}), giving the simple solution 
\begin{align}
\XL = \mathcal{A}^{-1/3} = \hat{\Omega}^{-2/3} \frac{1-e}{(1+e)^{1/3}} \frac{a}{r} \equiv \XLz(e,\hat{\Omega}) \frac{a}{r},
\end{align}
where we defined the function
\begin{align}
\XLz(e,\hat{\Omega})\equiv  \hat{\Omega}^{-2/3} \frac{1-e}{(1+e)^{1/3}}.
\end{align}

The magnitude of $\ve{r}_\Aa$, $r_\Aa$, could be the radius of the accretor star, or, if present, the size of the accretion disk. Generally, $r_\Aa$ could vary along the orbit. However, for simplicity, we assume that $r_\Aa$ is constant. 

In summary, for the ejection/accretion locations we assume
\begin{align}
\label{eq:rAmodel}
\rA = \left \{
\begin{array}{cc}
\displaystyle \left [ r(t) \XLz(q) - q r_\Aa \right ] \unit{r}(t), & \text{(Case 1)} \\ \\
\displaystyle \left [ a \XLz(e,\hat{\Omega}) - q r_\Aa \right ] \unit{r}(t). & \text{(Case 2)}
\end{array} \right.
\end{align}
Here, under the adiabatic approximation, $e$, $q$, $\hat{\Omega}$, and $r_\Aa$ are taken to be constant along one orbit when orbit-averaging (see \S\,\ref{sect:model:av}).

\subsection{Model for the mass transfer rate}
\label{sect:model:rate}
In order to proceed to find the orbit-averaged rates of change of the orbital elements implied by the simplified equations of motion (equation~\ref{eq:EOM_simple}), we need to specify the mass transfer rate as a function of orbital phase, i.e., $\dot{M}_\d=\dot{M}_\d(t)$. We recall from the Introduction that \citet{2007ApJ...667.1170S} and \citet{2016ApJ...825...71D} assumed instantaneous transfer at periapsis, i.e., a delta function $\dot{M}_\d \propto \delta(\theta)$. Here, we assume a more general model that resembles a delta function at very high eccentricity, but is also well defined for lower eccentricities, in particular, for circular orbits in which mass transfer is expected to occur continuously during the orbit. 

\begin{figure}
  \center
  \includegraphics[width=0.52\textwidth,trim=8mm 0mm 0mm 0mm]{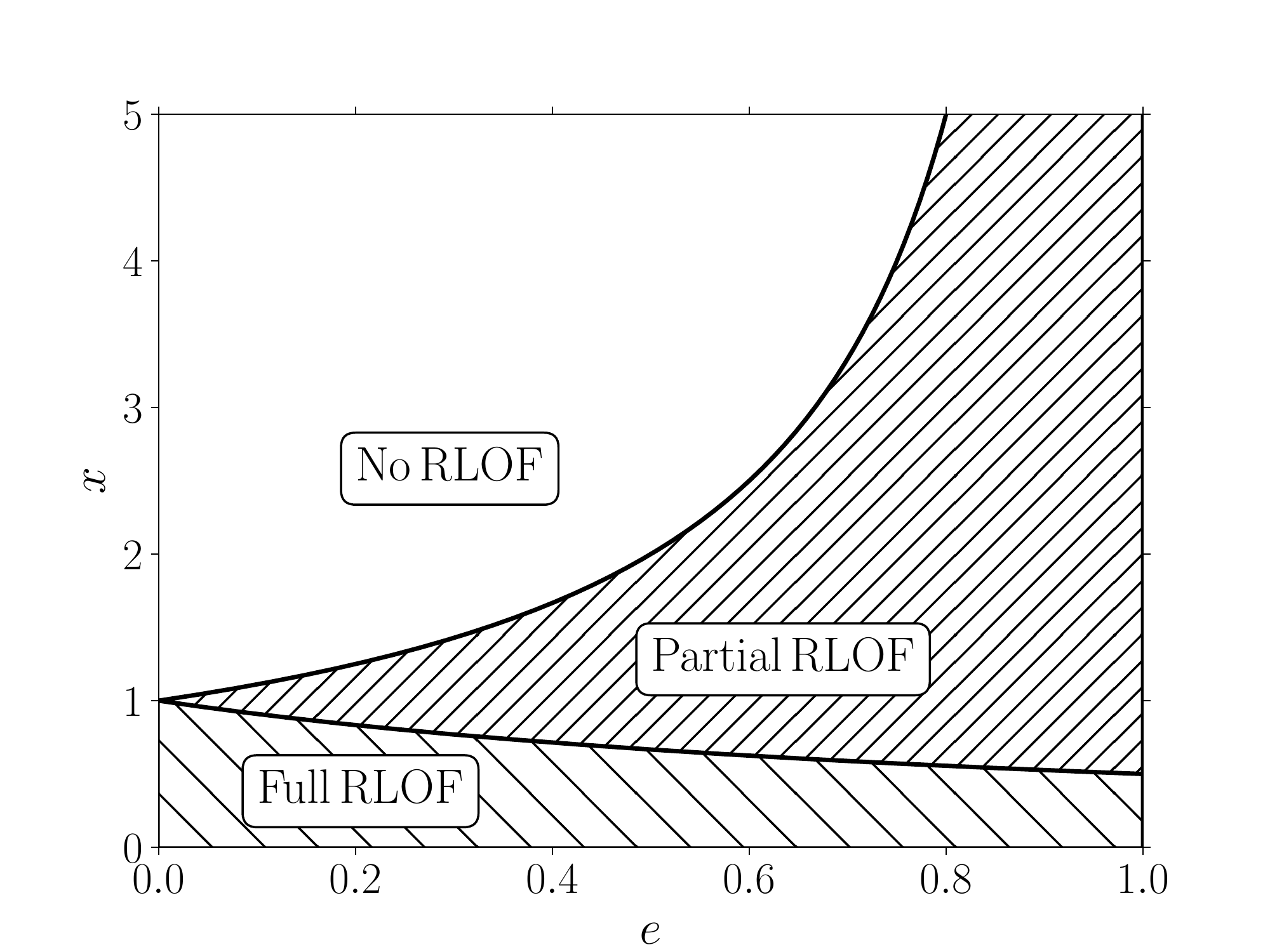}
  \caption{ Graphical representation of the regimes of mass transfer in our model (see \S\,\ref{sect:model:rate}) with the boundaries between the regimes plotted in the $(e,x)$ plane. The hatched regions correspond to RLOF (partial or full).}
\label{fig:RLOF}
\end{figure}

We assume that the instantaneous Roche lobe radius of the donor is given by
\begin{align}
R_\mathrm{L}(t) = \frac{R_\mathrm{L}^\mathrm{c}}{a} r(t),
\end{align}
where $r(t)$ is the instantaneous orbital separation, and
\begin{align}
\frac{R_\mathrm{L}^\mathrm{c}}{a} = \frac{0.49 \, q^{2/3}}{0.6 \, q^{2/3} + \ln\left (1 + q^{1/3} \right )}
\end{align}
is a fit to the Roche lobe radius in a circular orbit, provided by \citet{1983ApJ...268..368E}. A more accurate but more complicated expression for $R_\mathrm{L}(t)$ is given by \citet{2007ApJ...660.1624S}. For the moment, let the semimajor axis $a$ and the eccentricity $e$ be fixed. We consider three cases with regard to RLOF.
\begin{enumerate}
\item $R < R_\mathrm{L}(t)$: the donor does not fill its Roche lobe during any orbital phase (`no RLOF'). With $\dot{\Md} = 0$, equation~(\ref{eq:EOM_simple}) trivially reduces to the unperturbed two-body problem, and the orbit remains unchanged ($\dot{a} = \dot{e}=0$).
\item $R \geq R_\mathrm{L}(t)$ for any orbital phase: the donor fills its Roche lobe during the entire orbit (`full RLOF'). This is the case if the orbit is (close to) circular. 
\item $R \geq R_\mathrm{L}(t)$ for a range of orbital phases: the donor fills its Roche lobe during part of the orbit (`partial RLOF'). This occurs if the orbit is (sufficiently) eccentric. 
\end{enumerate}
Defining
\begin{align}
\label{eq:x_def}
x \equiv  \frac{R_\mathrm{L}^\mathrm{c}}{R} = \frac{a}{R} \frac{0.49 \, q^{2/3}}{0.6 \, q^{2/3} + \ln\left (1 + q^{1/3} \right )},
\end{align}
these conditions can be written as
\begin{enumerate}
\item $\displaystyle x > \frac{1}{1-e}$; \quad (no RLOF)
\item $\displaystyle x \leq \frac{1}{1+e}$; \quad (full RLOF)
\item $\displaystyle \frac{1}{1+e} < x \leq \frac{1}{1-e}$. \quad (partial RLOF)
\end{enumerate}
The three cases are illustrated graphically in \F\,\ref{fig:RLOF} with the boundaries plotted in the $(e,x)$ plane, and with the hatched regions corresponding to RLOF (partial or full). For $e=0$, the three cases reduce to two cases consistent with the standard picture of RLOF in circular orbits: no RLOF if $x > 1$, and (full) RLOF if $x \leq 1$. In the circular case, no RLOF is possible if $x>1$ (i.e., $R<R_\mathrm{L}^\mathrm{c}$). In the eccentric case, however, partial RLOF is possible provided that $e$ is large enough, i.e., $e>1-1/x$. 

The range of orbital phases for which RLOF occurs in case (3) is given by $-\EA_0<\EA<\EA_0$, where $\EA$ is the eccentric anomaly (of course, the mean and true anomalies are equally valid variables to describe this range). Using the canonical relation 
\begin{align}
r(\EA) = a(1-e\cos \EA),
\end{align}
$\EA_0$ is given by 
\begin{align}
\label{eq:E_0}
\cos \EA_0 = \frac{1}{e} \left (1 - \frac{1}{x} \right ).
\end{align}
Note that, strictly speaking, case (2) can be considered as a special case of case (3), i.e., case (3) reduces to case (2) if $\EA_0 = \pi$. 

One approach to model the phase-dependent mass transfer rate might be to assume a step function that is zero in case (1), and constant in cases (2) and (3) in the range $-\EA_0<\EA<\EA_0$. However, a physically more realistic model should take into account the known property that the mass transfer rate is highly sensitive to the `radius excess' $\Delta R \equiv R-R_\mathrm{L}$, i.e., the degree at which the donor overflows its Roche lobe (e.g., \citealt{2011ApJ...726...66L}). Therefore, the mass transfer rate is expected to be higher closer to periapsis, where the (instantaneous) Roche lobe radius is smaller. 

For a donor with an atmospheric scale height $H_\mathrm{P}$, \citet{1988A&A...202...93R} derived that the mass transfer rate should be
\begin{align}
\label{eq:M_d_exp}
\dot{\Md} \propto \exp \left ( \frac{R-R_\mathrm{L}}{H_\mathrm{P}} \right ),
\end{align}
where $H_\mathrm{P}$ is the pressure scale height of the donor's atmosphere. Equation~(\ref{eq:M_d_exp}) states that the transfer rate increases exponentially with $\Delta R$. Another result can be obtained by assuming a polytropic equation of state for the donor coupled with Bernoulli's equation, giving (\citealt{1972AcA....22...73P,1987MNRAS.229..383E} \footnote{See also section 7.1, pages 8-9, of lecture notes on binary star evolution by Onno Pols, which are downloadable at the url \href{https://www.astro.ru.nl/~onnop/education/binaries_utrecht_notes/Binaries_ch6-8.pdf}{https://www.astro.ru.nl/$\sim$onnop/education/binaries\_utrecht\_notes/Binaries\_ch6-8.pdf}.})
\begin{align}
\label{eq:M_d_poly}
\dot{\Md} \propto \left ( \frac{R-R_\mathrm{L}}{R} \right )^{n_\mathrm{p}+3/2},
\end{align}
where $n_\mathrm{p}$ is the polytropic index. 

Ideally, one should consider both cases: equation~(\ref{eq:M_d_exp}), and equation (\ref{eq:M_d_poly}) with a general $n_\mathrm{p}$. However, with these functional relations for $\dot{M}_\d$, we were unable to analytically compute the resulting integrals required to derive the orbit-averaged equations of motion (see \S\,\ref{sect:model:av} below). Fortunately, however, we were able to do so in the specific case of equation~(\ref{eq:M_d_poly}) with $n_\mathrm{p}=3/2$, which is a reasonable approximation for a wide range of objects such as convective stars and low-mass white dwarfs (e.g., \citealt{1939isss.book.....C}), and gas giant planets (e.g., \citealt{2015MNRAS.452.1375W}). 

Therefore, we choose to adopt the power-law dependence of $\dot{M}_\d$ with $n_\mathrm{p}=3/2$, i.e., in our model, we set
\begin{align}
\label{eq:Md_tau_zero}
\dot{M}_\d = \dot{M}_{\d,\,0} \left ( \frac{R-R_\mathrm{L}}{R} \right )^3 = \dot{\Md}_{,0} \left [ 1 - x (1-e \cos \EA) \right]^3,
\end{align}
where $\dot{M}_{\d,\,0}$ is a phase-independent mass transfer rate. We note that equation~(\ref{eq:Md_tau_zero}) states that $\dot{M}_\d$ is a strong function of $\Delta R$, similarly to equation~(\ref{eq:M_d_exp}). Also, we emphasize that equation~(\ref{eq:Md_tau_zero}) only gives a qualitative description of the mass transfer rate; other choices might be equally valid. 

\begin{figure}
  \center
  \includegraphics[width=0.52\textwidth,trim=5mm 0mm 0mm 0mm]{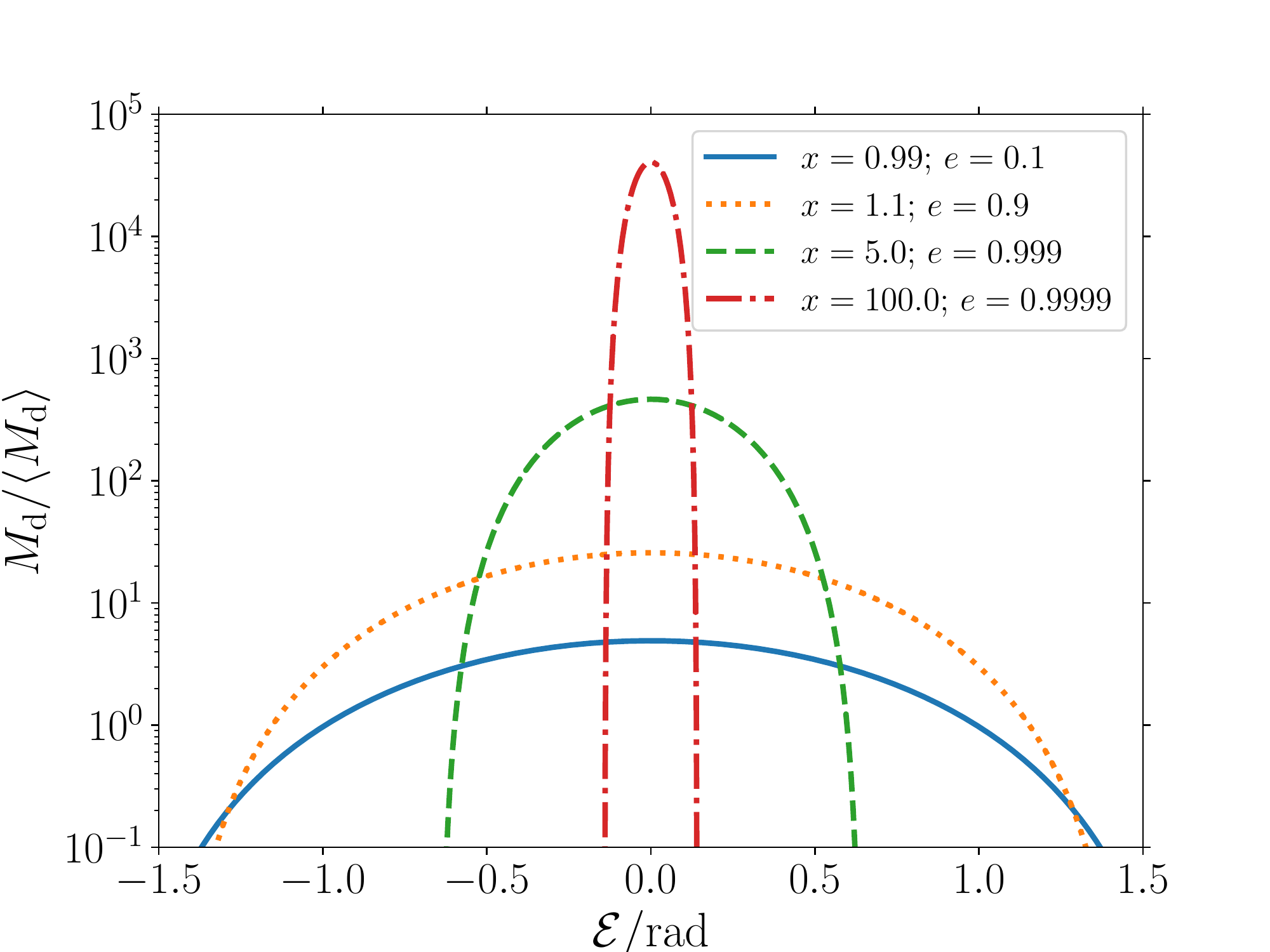}
  \caption{Mass transfer rate $\dot{M}_\d$ in our model (equation~\ref{eq:Md_tau_zero}), normalized to $\langle \dot{M}_\d\rangle$, plotted as a function of orbital phase, $\EA$, for different values of $x$ and $e$ (refer to the legend). }
\label{fig:M_d_dot}
\end{figure}

Below, we relate $\dot{M}_{\d,\,0}$ to the orbit-averaged mass transfer rate $\langle \dot{M}_\d\rangle$; we assume that $\langle \dot{M}_\d\rangle$ is known. For illustration, we plot in \F\,\ref{fig:M_d_dot} the mass transfer rate $\dot{M}_\d$ normalized to $\langle \dot{M}_\d\rangle$ as a function of the eccentric anomaly, $\EA$, for different values of $x$ and $e$. For low eccentricity and $x$ close to zero, the mass transfer rate is nearly constant. As $x$ and $e$ increase, the mass transfer rate becomes increasingly peaked around $\EA=0$ (with a decreasing width, and increasing height). 

From the equations of motion, equation~(\ref{eq:EOM}), it is clear that the double time derivative $\ddot{M}_\d$ should also be specified. From equation~(\ref{eq:Md_tau_zero}), we directly obtain $\ddot{M}_\d$ assuming that $\dot{\Md}_{,0}$, $x$, and $e$ are constant during the orbit, i.e.,
\begin{align}
\label{eq:Mdd_tau_zero}
\ddot{M}_\d = - 3 n x e \dot{M}_{\d,\,0} \left [ 1 - x (1-e \cos \EA) \right]^2 \frac{\sin\EA}{1-e\cos\EA},
\end{align}
where we used the canonical relation $\dot{\EA} = n a/r$.

As a further sophistication, we introduce a time lag $\tau$ (with dimensions of time) to take into account the known phenomenon that the mass transfer rate is not symmetric around the apsidal line, but peaks just after periapsis (e.g., \citealt{2016MNRAS.455..462V}). In this case, we evaluate the Roche lobe radius in the expression for the mass transfer rate at $t-\tau$, i.e.,
\begin{align}
\label{eq:Md_tau_nonzero}
\nonumber \dot{M}_\d &= \dot{M}_{\d,\,0} \left ( \frac{R-R_\mathrm{L}(t-\tau)}{R} \right )^3 \\
&= \dot{\Md}_{,0} \left [ 1 - x \left \{ 1-e \cos (\EA-\EA_\tau) \right \} \right]^3,
\end{align}
where $\EA_\tau$ is the eccentric anomaly corresponding to the time interval $\Delta t =\tau$ as measured from periapsis. The quantity $\EA_\tau$ is a function of $\tau$, the mean motion $n$, and the eccentricity, and is found using the Kepler equation, i.e.,
\begin{align}
\label{eq:E_tau}
n \tau = \EA_\tau - e \sin \EA_\tau.
\end{align}
For small $\tau$ ($n \tau \ll 1$), equation~(\ref{eq:E_tau}) yields the approximate solution $\EA_\tau \approx n\tau/(1-e)$. Below, we distinguish between the cases $\tau=0$, and $\tau\neq0$. It turns out that the expressions associated with the ejection/accretion radii in the orbit-averaged equations become excessively complicated if $\tau\neq 0$ (i.e., with hundreds of terms appearing). Therefore, when presenting the equations below in \S\,\ref{sect:model:av}, in the case $\tau\neq0$ we include only the terms that appear if $\ve{r}_\Ad = \ve{r}_\Aa = \ve{0}$.

\subsection{Orbit-averaged equations of motion}
\label{sect:model:av}
\subsubsection{Preliminaries}
\label{sect:model:av:pre}
For completeness, we first give the standard equations that describe the evolution of the orbital elements of the binary (see, e.g., Appendix C of \citealt{2006epbm.book.....E}). Write the relative acceleration as
\begin{align}
\frac{\mathrm{d}^2 \ve{r}}{\mathrm{d} t^2} &=  -\frac{GM}{r^3} \ve{r} + \ve{f},
\end{align}
where $\ve{f}$ is the perturbing acceleration (force per unit mass). In our case, the perturbing acceleration is
\begin{align}
\label{eq:f_simple}
\ve{f} = - \frac{\dot{M}_\d}{M_\d} \left (1-q \right )\dot{\ve{r}} - \frac{\dot{M}_\d}{M_\d} \bm{\omega}_\orb \times \rA -  \frac{\ddot{M}_\d}{M_\d} \rA,
\end{align}
where $\dot{\ve{r}}$, $\bm{\omega}_\orb = n \sqrt{1-e^2} (a/r)^2 \, \unit{h}$, $\dot{M}_\d$, $\ddot{M}_\d$, and $\rA$ are all functions of the orbital phase (see equations~\ref{eq:Md_tau_zero}, \ref{eq:Mdd_tau_zero} and \ref{eq:rAmodel} for the latter three terms, respectively), whereas $M_\d$ and $q$ are assumed to be phase independent (adiabatic regime).

The perturbation $\ve{f}$ gives rise to a change in the specific orbital energy $\Ec = \frac{1}{2} (\dot{\ve{r}} \cdot \dot{\ve{r}}) - G M /r$ given by
\begin{align}
\label{eq:dotE}
\dot{\Ec} = - \Ec \,(\dot{a}/a) = \dot{\ve{r}} \cdot \ve{f}.
\end{align}
Furthermore, the change in the eccentricity vector, 
\begin{align}
\ve{e} = \frac{1}{GM} \dot{\ve{r}} \times \ve{h} - \unit{r} = \frac{1}{GM} \left [ \ve{r} (\dot{\ve{r}} \cdot \dot{\ve{r}}) - \dot{\ve{r}} (\ve{r} \cdot \dot{\ve{r}} ) \right ] - \unit{r},
\end{align}
where $\ve{h} \equiv \ve{r} \times \dot{\ve{r}}$ is the specific angular momentum vector, is given by
\begin{align}
\label{eq:dote}
\dot{\ve{e}} = \frac{1}{GM} \left [ 2 \ve{r} \left(\dot{\ve{r}} \cdot \ve{f} \right ) - \ve{f} \left ( \ve{r} \cdot \dot{\ve{r}} \right ) - \dot{\ve{r}} \left ( \ve{r} \cdot \ve{f} \right ) \right ].
\end{align}
The specific angular momentum vector changes as a result of the torque, i.e.,
\begin{align}
\dot{\ve{h}} = \ve{r} \times \ve{f}.
\end{align}
It is important to note that equations~(\ref{eq:dotE}) and (\ref{eq:dote}) assume that the total mass $M$ is constant. This is the case for conservative mass transfer ($\dot{M} = 0$), but not necessarily for nonconservative mass transfer (the latter is beyond the scope of this paper).

The first term of equation~(\ref{eq:f_simple}) is proportional to the relative velocity, $\dot{\ve{r}}$. For a perturbation of the form $\tilde{\ve{f}} = C \,\dot{\ve{r}}$, where $C$ is a scalar quantity that can depend on time/orbital phase, the above expressions imply
\begin{subequations}
\begin{align}
\label{eq:adotgen} \frac{\dot{a}}{a} &= 2C \frac{1+e \cos \EA}{1-e \cos \EA}; \\
\label{eq:edotgen} \dot{\ve{e}} &= \frac{2C}{1-e \cos\EA} \left [ \left (1-e^2 \right ) \cos\EA \, \unit{e} + \sqrt{1-e^2}  \sin\EA \, \unit{q} \right ],
\end{align}
\end{subequations}
where $\unit{q} \equiv \unit{h} \times \unit{e}$. Equation~(\ref{eq:edotgen}) implies a scalar eccentricity change of
\begin{align}
\dot{e} = \unit{e} \cdot \dot{\ve{e}} = 2C \frac{1-e^2}{1-e \cos\EA} \cos \EA,
\end{align}
and the argument of periapsis $\omega$ changes according to
\begin{align}
\dot{\omega} = \frac{\unit{q} \cdot \dot{\ve{e}}}{e} = \frac{2C}{e} \frac{\sqrt{1-e^2}}{1-e \cos\EA} \sin \EA.
\end{align}
Under the influence of $\tilde{\ve{f}}=C \,\dot{\ve{r}}$, the inclination $i$ and longitude of the ascending node $\Omega$ remain constant, since $\dot{\ve{h}} = C \, \ve{r} \times \dot{\ve{r}} = C \,\ve{h}$, implying that $\ve{h}$ does not change its direction. 

More generally, for any phase-dependent perturbation $\tilde{\ve{f}}$, the orbital elements change according to
\begin{subequations}
\begin{align}
\frac{\dot{a}}{a} &= \frac{2na^2}{GM(1-e\cos\EA)} \left [ -\left (\unit{e} \cdot \tilde{\ve{f}} \right ) \sin \EA + \sqrt{1-e^2}\left(\unit{q} \cdot \tilde{\ve{f}} \right ) \cos \EA \right ]; \\
\nonumber \dot{e} &= \frac{na^2}{GM(1-e\cos\EA)} \left [ - \left(1-e^2\right) \left (\unit{e} \cdot \tilde{\ve{f}} \right ) \sin \EA \cos\EA \right. \\
&\quad \left. + \sqrt{1-e^2} \left(\unit{q} \cdot \tilde{\ve{f}} \right )\left (1-2e\cos \EA + \cos^2\EA \right ) \right ]; \\
\nonumber \dot{\omega} &= \frac{na^2}{GMe(1-e\cos\EA)} \left [ \sqrt{1-e^2} \left ( \unit{e} \cdot \tilde{\ve{f}} \right ) \left (-2 + e \cos \EA + \cos^2\EA\right ) \right. \\
&\quad \left. + \left(\unit{q} \cdot \tilde{\ve{f}} \right ) \left (\cos \EA - e \right ) \sin\EA \right ].
\end{align}
\end{subequations}
We assume that $\rA$ in the second and third terms of equation~(\ref{eq:f_simple}) is directed along $\unit{r}$. Therefore, the second term in equation~(\ref{eq:f_simple}) is $\propto \unit{h} \times \unit{r}$, implying that the associated $\dot{\ve{h}} \propto \ve{r} \times \left (\unit{h} \times \unit{r} \right) = r \, \unit{h}$, i.e., $\ve{h}$ does not change direction as a result of the second term. Lastly, due to the third term in equation~(\ref{eq:f_simple}) $\dot{\ve{h}} \propto \ve{r} \times \unit{r} = \ve{0}$, i.e., $\ve{h}$ does not change at all as a result of the third term. We conclude that, for our assumed perturbation in equation~(\ref{eq:f_simple}), $i$ and $\Omega$ remain constant.

Next, we orbit-average the remaining nontrivial equations for $\dot{a}$, $\dot{e}$ and $\dot{\omega}$. 
We define orbit-averaged quantities in the usual way and formulated in terms of the eccentric anomaly, i.e.,
\begin{align}
\label{eq:av}
\langle (...) \rangle = \frac{1}{2\pi} \int_{-\pi}^{\pi} \, \mathrm{d} \EA \, \left ( \frac{r}{a} \right ) \, ( ... ),
\end{align}
where $(...)$ denotes the quantity to be averaged. In equation~(\ref{eq:av}), we assume that the orbital elements (most importantly, $a$ and $e$) are constant during the orbit (adiabatic approximation). In our model, the mass transfer rate is zero for $\EA<-\EA_0$, and $\EA>\EA_0$. Therefore, in practice, the range of the integral in equation~(\ref{eq:av}) is taken to be $-\EA_0<\EA<\EA_0$. Note that $\EA_0$ is a function of $e$ and $x$ (see equation~\ref{eq:E_0}). 

\paragraph{Normalization} As mentioned in \S\,\ref{sect:model:rate}, we relate the quantity $\dot{\Md}_{,0}$ in equation~(\ref{eq:Md_tau_zero}) to the orbit-averaged mass transfer rate $\langle \dot{M}_\d \rangle$, and assume that $\langle \dot{M}_\d \rangle$ is known. For $\tau=0$, equations~(\ref{eq:Md_tau_zero}) and (\ref{eq:av}) imply
\begin{align}
\langle \dot{M}_\d \rangle = \dot{M}_{\d,\,0}  f_{\dot{M}} (e,x),
\end{align}
where the dimensionless function $f_{\dot{M}} (e,x)$ is given explicitly by equation~(\ref{eq:f_M}) in Appendix~\ref{app:funcs}. When $\tau\neq0$, the dimensionless function should be replaced with $f_{\dot{M}} (e,x,\EA_\tau)$, which is given explicitly by equation~(\ref{eq:f_M_tau}) in Appendix~\ref{app:funcs}.

We remark that $\langle \dot{M}_\d \rangle$ is assumed to be known and constant during the orbit. However, $\langle \dot{M}_\d \rangle$ can change owing to the changing structure of the donor as a result of mass transfer, and/or owing to stellar evolution. For simplicity and to separate the orbital evolution from the donor's structure and stellar evolution, we assume that $\langle \dot{M}_\d \rangle$ is constant in our applications (\S s~\ref{sect:appl_bin} and \ref{sect:appl_triple}). However, in other situations, e.g., when modeling the long-term evolution in population synthesis studies, $\langle \dot{M}_\d \rangle$ should be calculated self-consistently and allowed to vary over timescales much longer than the orbital timescale.

\subsubsection{Secular change of the orbital elements with $\tau=0$ --- case (1): negligible donor spin}
\label{sect:model:av:results_case_1}
Setting $\tau=0$ and in case (1) of \S\,\ref{sect:model:rA} for $\rAd$ (negligible donor spin), the orbit-averaged equations of motion are given by
\begin{subequations}
\label{eq:av_EOM_case_1}
\begin{align}
\nonumber \displaystyle\frac{\langle \dot{a} \rangle}{a} &= \displaystyle -\frac{2 \langle \dot{\Md} \rangle}{\Md} \frac{1}{f_{\dot{M}}(e,x) } \Biggl [ (1-q) f_a(e,x) + \XLz(q) g_a(e,x)\\
&\qquad \qquad \displaystyle  - q \frac{r_\Aa}{a} h_a(e,x) \Biggl ]; \\
\nonumber \displaystyle \langle \dot{e} \rangle &= \displaystyle -\frac{2 \langle \dot{\Md} \rangle}{\Md} \frac{1}{f_{\dot{M}}(e,x) } \Biggl [ (1-q) f_e(e,x) + \XLz(q) g_e(e,x)\\
&\qquad \qquad \displaystyle  - q \frac{r_\Aa}{a} h_e(e,x) \Biggl ]; \\
\displaystyle \langle \dot{\omega} \rangle &= 0.
\end{align}
\end{subequations}
The dimensionless quantities $g_a(e,x)$, $h_a(e,x)$, $g_e(e,x)$, and $h_e(e,x)$ are closed-form analytic functions given explicitly in Appendix~\ref{app:funcs}. In presenting these functions, we keep the explicit dependence on $\EA_0=\EA_0(e,x)$ (see equation~\ref{eq:E_0}) which appears through the integration limits. In our practical implementation, $\EA_0$ is replaced using equation~(\ref{eq:E_0}) instead of carrying out this replacement analytically. This approach turns out to be numerically favorable, since situations can otherwise occur in which the ODE integrator evaluates the functions at illegal combinations of $e$ and $x$ (i.e., such that $\cos \EA_0$ does not lie between -1 and 1). 

Note that there is no secular change of the argument of periapsis, $\omega$. This is no longer the case when $\tau\neq0$ (see \S\,\ref{sect:model:av:results_nonzero_tau}).

\subsubsection{Secular change of the orbital elements with $\tau=0$ --- case (2): large mass ratio}
\label{sect:model:av:results_case_2}
Setting $\tau=0$ and in case (2) of \S\,\ref{sect:model:rA} for $\rAd$ (large mass ratio $q$), the orbit-averaged equations of motion are given by
\begin{subequations}
\begin{align}
\label{eq:av_EOM_case_2}
\nonumber \displaystyle\frac{\langle \dot{a} \rangle}{a} &= \displaystyle -\frac{2 \langle \dot{\Md} \rangle}{\Md} \frac{1}{f_{\dot{M}}(e,x) } \Biggl [ (1-q) f_a(e,x) \\
&\qquad \qquad \displaystyle + \left ( \XLz(e,\hat{\Omega}) - q \frac{r_\Aa}{a} \right ) h_a(e,x) \Biggl ]; \\
\nonumber \displaystyle \langle \dot{e} \rangle &= \displaystyle -\frac{2 \langle \dot{\Md} \rangle}{\Md} \frac{1}{f_{\dot{M}}(e,x) } \Biggl [ (1-q) f_e(e,x) \\
&\qquad \qquad \displaystyle + \left ( \XLz(e,\hat{\Omega}) - q \frac{r_\Aa}{a} \right ) h_e(e,x) \Biggl ]; \\
\displaystyle \langle \dot{\omega} \rangle &= 0.
\end{align}
\end{subequations}

\subsubsection{Secular change of the orbital elements with $\tau\neq0$ and zero ejection/accretion radii}
\label{sect:model:av:results_nonzero_tau}
As mentioned in \S\,\ref{sect:model:rate}, the terms associated with $\ve{r}_\Ad$ and $\ve{r}_\Aa$ are excessively complicated if $\tau\neq0$. Therefore, we here restrict to the secular orbital evolution associated with the term $\propto \dot{\ve{r}}$ in equation~(\ref{eq:f_simple}) only, i.e., setting $\rAd=\rAa=\ve{0}$. In this case, the orbit-averaged equations of motion are given by 

\begin{subequations}
\label{eq:av_EOM_tau_nonzero}
\begin{align}
\displaystyle\frac{\langle \dot{a} \rangle}{a} &= \displaystyle -\frac{2 \langle \dot{\Md} \rangle}{\Md}(1-q) \frac{f_a(e,x,\EA_\tau)}{f_{\dot{M}}(e,x) }; \\
\displaystyle \langle \dot{e} \rangle &= \displaystyle -\frac{2 \langle \dot{\Md} \rangle}{\Md}(1-q) \frac{f_e(e,x,\EA_\tau)}{f_{\dot{M}}(e,x) }; \\
\displaystyle \langle \dot{\omega} \rangle &= \displaystyle -\frac{2 \langle \dot{\Md} \rangle}{\Md} (1-q)\frac{f_\omega(e,x,\EA_\tau)}{f_{\dot{M}}(e,x) } .
\end{align}
\end{subequations}
The functions $f_a(e,x,\EA_\tau)$, $f_e(e,x,\EA_\tau)$, and $f_\omega(e,x,\EA_\tau)$ are given explicitly in Appendix~\ref{app:funcs}. We recall that $\EA_\tau$ is a function of $\tau$, $n$, and $e$ (see equation~\ref{eq:E_tau}).

\subsection{Properties of the orbit-averaged equations of motion}
\label{sect:model:prop}
Although the general expressions for the dimensionless functions in the orbit-averaged equations of motion are cumbersome (see Appendix\,\ref{app:funcs}), some insight can be gained in limiting cases. Here, we consider the functions $f_a(e,x,\EA_\tau)$, $f_e(e,x,\EA_\tau)$, and $f_\omega(e,x,\EA_\tau)$, all normalized to $f_{\dot{M}}(e,x,\EA_\tau)$. 
 
In the limit of $\EA_0\rightarrow \pi$, RLOF occurs during the entire orbit. In this case, setting $\rAd$ and $\rAa$ to zero and to second order in the eccentricity, the equations of motion are given by (assuming $x\neq1$)
\begin{subequations}
\label{eq:av_EOM_tau_nonzero_limit}
\begin{align}
\label{eq:av_EOM_tau_nonzero_limit_a} \displaystyle\frac{\langle \dot{a} \rangle}{a} &= \displaystyle -\frac{2 \langle \dot{\Md} \rangle}{\Md}(1-q) \Biggl [1-\frac{3 e^2 x \cos (\EA_\tau)}{x-1}+O\left(e^3\right) \Biggl ]; \\
\label{eq:av_EOM_tau_nonzero_limit_e}\displaystyle \langle \dot{e} \rangle &= \displaystyle -\frac{2 \langle \dot{\Md} \rangle}{\Md}(1-q) \Biggl [\frac{3}{2} e \cos \EA_\tau \frac{x}{1-x}+O\left(e^3\right) \Biggl ]; \\
\label{eq:av_EOM_tau_nonzero_limit_omega} \nonumber \displaystyle \langle \dot{\omega} \rangle &= \displaystyle -\frac{2 \langle \dot{\Md} \rangle}{\Md} (1-q) \sin \EA_\tau \Biggl [\frac{3}{2} \frac{x}{1-x} + \frac{3}{8} e^2 x \frac{1}{
   (x-1)^3}  \\
& \times \biggl \{ 2+ 6 (x-1) x \cos \EA_\tau+7 x^2-4 x+2 \biggl \} +O\left(e^3\right) \Biggl ].
\end{align}
\end{subequations}
As the orbit circularizes ($e\rightarrow0$), equation~(\ref{eq:av_EOM_tau_nonzero_limit_a}) reduces to the canonical relation for conservative mass transfer in circular orbits, equation~(\ref{eq:a_dot_can}). Furthermore, equation (\ref{eq:av_EOM_tau_nonzero_limit_e}) states that, as $e\rightarrow0$, $\langle \dot{e} \rangle \propto e$, i.e., $e$ decays exponentially to zero. In contrast, according to the delta function model of \citet{2007ApJ...667.1170S} and \citet{2016ApJ...825...71D}, for low eccentricities $\langle \dot{e} \rangle \propto \sqrt{1-e^2}(1-e) \approx 1-e$, i.e., $\langle \dot{e} \rangle$ is nonzero at $e=0$, and this can lead to undesirable properties (see \S\,\ref{sect:appl_bin} below for explicit examples). This is a consequence of the assumption of a delta function mass transfer rate at periapsis, whereas periapsis is not defined for circular orbits. 

Lastly, note that $ \langle \dot{\omega} \rangle\propto \sin \EA_\tau$ vanishes for $\EA_\tau=0$ ($\tau=0$), i.e., the orbit only secularly precesses if mass is transferred asymmetrically with respect to periapsis.

\subsection{Numerical implementation}
\label{sect:model:num}
To numerically integrate the orbit-averaged equations of motion as presented in \S s\,\ref{sect:model:av:results_case_1}, \ref{sect:model:av:results_case_2}, and \ref{sect:model:av:results_nonzero_tau}, we implemented them into a code called \textsc{emt} (Eccentric Mass Transfer) which is freely available at \href{https://github.com/hamers/emt}{https://github.com/hamers/emt}. The code is an easy-to-use and standalone \textsc{Python} script (using the standard \textsc{NumPy} and \textsc{SciPy} libraries). The functions, as given in Appendix~\ref{app:funcs}, are implemented in \textsc{C} and interfaced with \textsc{Python} using \textsc{ctypes}. These functions can therefore be easily ported to other codes such as binary or higher-order multiplicity population synthesis codes, which are typically written in \textsc{C} or similar languages. In addition, we implemented the secular equations of motion associated with a third body, using the standard expressions to quadrupole and octupole order (e.g., \citealt{1968AJ.....73..190H,2000ApJ...535..385F,2013MNRAS.431.2155N}). We also included the first-order post-Newtonian (PN) terms in the inner orbit that give rise to orbital precession. However, we did not include higher-order PN terms or tidal effects (tidal dissipation and tidal bulges).

Below, in \S s\,\ref{sect:appl_bin} and \ref{sect:appl_triple}, we use the \textsc{emt} code to investigate mass transfer in binary and triple systems, respectively.

\section{Applications: mass transfer in circular and eccentric binaries}
\label{sect:appl_bin}
We apply our model to a number of systems using the numerical implementation described in \S\,\ref{sect:model:num}. In \S\,\ref{sect:appl_bin}, we restrict to the case of isolated binaries and concentrate on comparing our results to those of \citet{2007ApJ...667.1170S} and \citet{2016ApJ...825...71D}. In \S\,\ref{sect:appl_triple}, we focus on hierarchical triple systems undergoing LK oscillations. 

\begin{figure}
  \center
  \includegraphics[width=0.52\textwidth,trim=8mm 0mm 0mm 0mm]{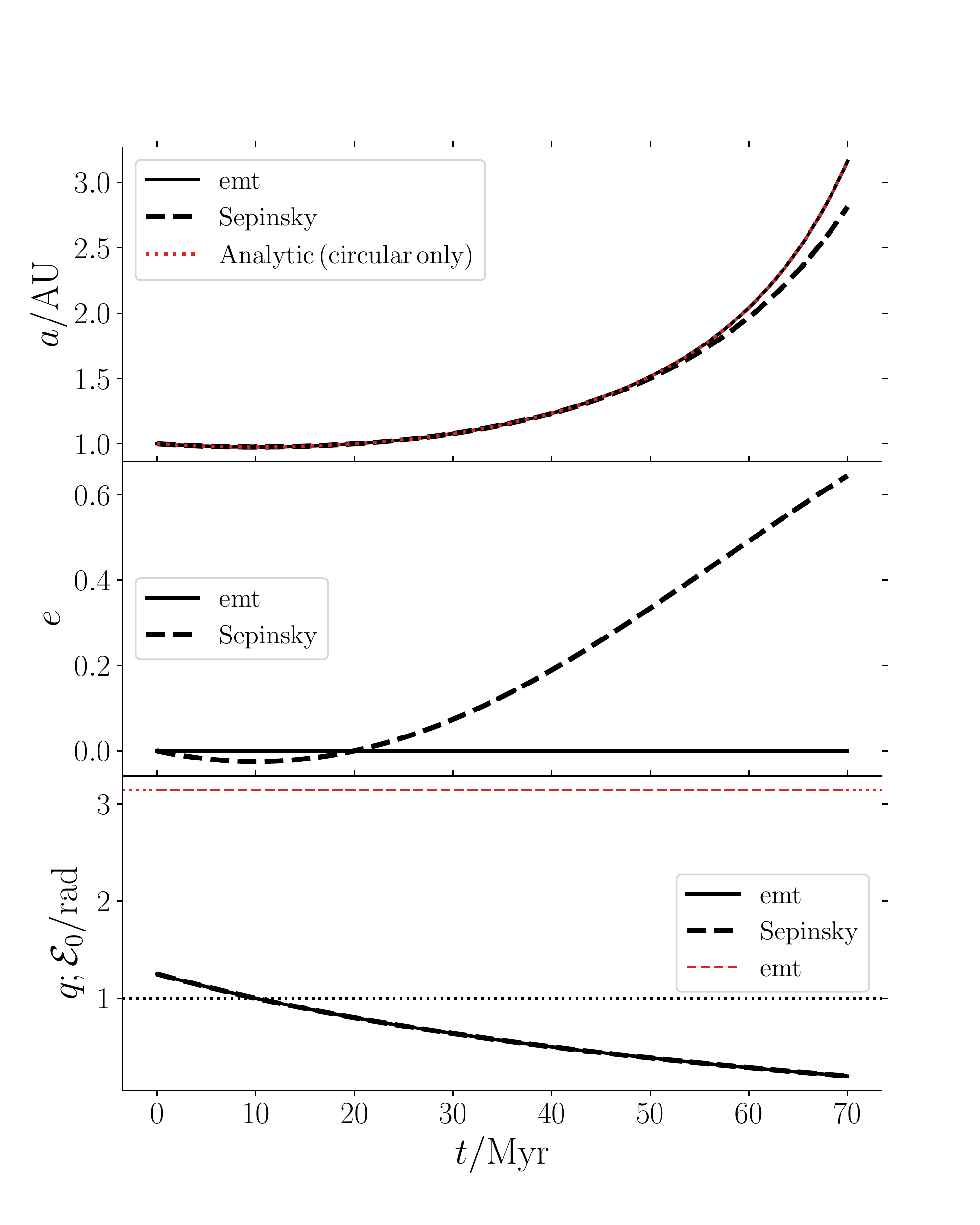}
  \caption{ Semimajor axis (top panel), eccentricity (middle panel), and mass ratio $q$ and $\EA_0$ (bottom panel) as a function of time for a circular binary, and setting $\rAd=\rAa=\ve{0}$ (see \S\,\ref{sect:appl_bin:circ} for the initial conditions). Results are shown according to our model (labeled `emt'; black solid lines), and the model from \citet{2007ApJ...667.1170S} and \citet{2016ApJ...825...71D} (labeled `Sepinsky; black dashed lines). In the bottom panel, the black (red) lines show $q$ ($\EA_0$). In the top panel, the canonical analytic expectation, $M_\d^2M_\a^2a$ is constant, is shown with the red dotted line. In the bottom panel, the black dotted line shows $q=1$, and the red dotted line shows $\EA_0=\pi$ (note that also $\EA_0=\pi$ in the `emt' model). }
\label{fig:binary_circular}
\end{figure}

\begin{figure}
  \center
  \includegraphics[width=0.52\textwidth,trim=8mm 0mm 0mm 0mm]{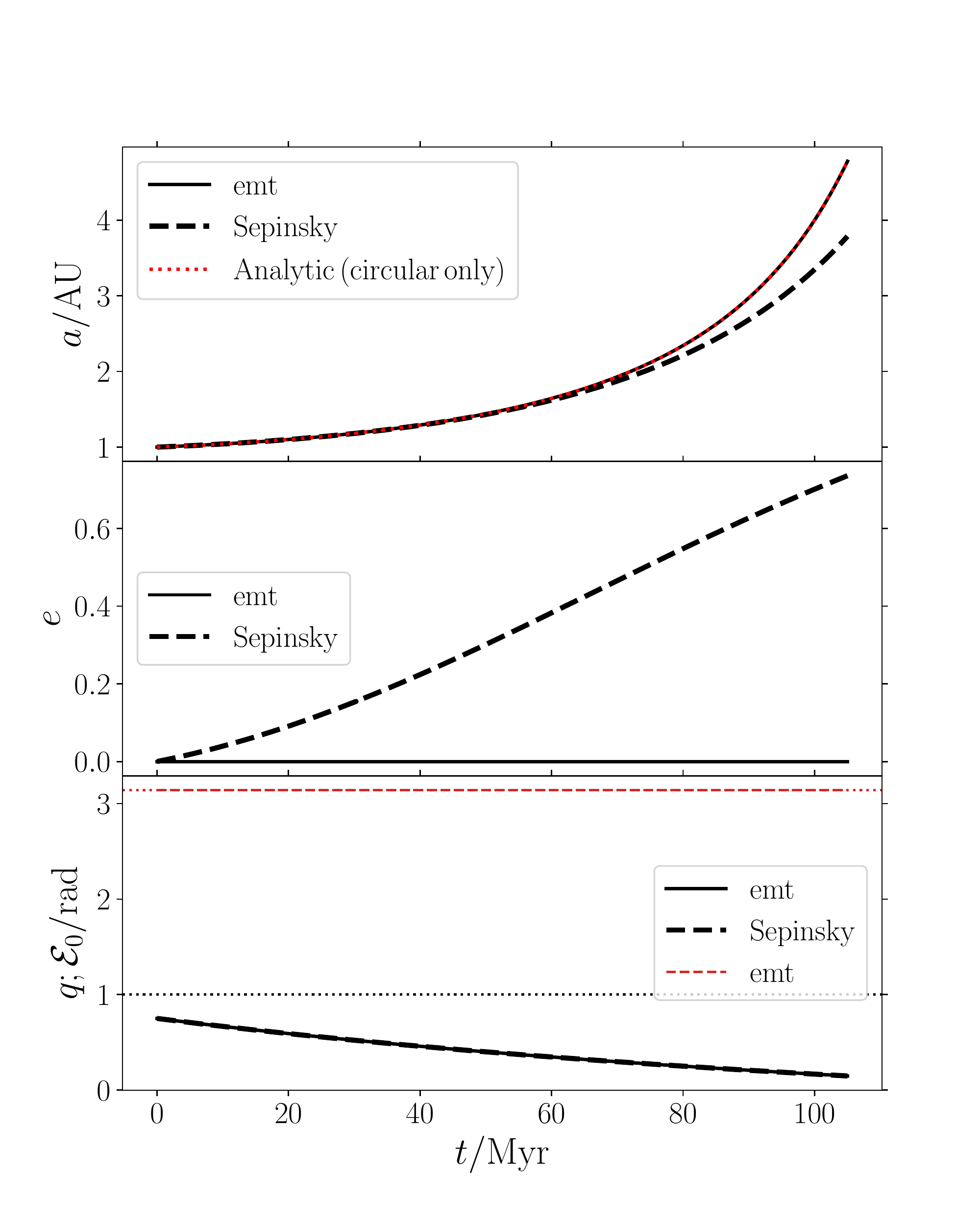}
  \caption{ Similar to \F\,\ref{fig:binary_circular}, except that now the initial mass ratio is $q=1.5/2.0<1$. }
\label{fig:binary_circular_low_q}
\end{figure}

\subsection{Circular orbit}
\label{sect:appl_bin:circ}
We first consider the simplest case of mass transfer in a circular binary, also setting $\rAa=\rAd=\ve{0}$ (point masses). As mentioned above, the expected orbital evolution in this case is described by equation~(\ref{eq:a_dot_can}). From the fact that, for a circular orbit and conservative mass transfer, $L_\orb^2 = \mu^2 GMa$ and $M$ are constant, it immediately follows that $M_\d^2M_\a^2a$ is conserved during mass transfer. 

In our numerical example, we set the initial parameters to $M_\d=1\,\msun$, $M_\a=0.8\,\msun$, $\langle \dot{M}_\d\rangle = -10^{-8}\,\msun\,\mathrm{yr^{-1}}$, $a=1\,\au$, and $R=1000\,\rsun$, i.e., RLOF triggered in a relatively tight binary when a solar-type primary evolves to a giant. We emphasize that our parameters are chosen for illustration purposes only, and may not be entirely realistic. Nonetheless, in this case the qualitative behavior of mass transfer is independent of the exact choice of parameters.

We show the evolution of the semimajor axis and eccentricity in the top and middle panels of \F\,\ref{fig:binary_circular}, respectively. The bottom panel shows the mass ratio $q$ and $\EA_0$ (if applicable) as a function of time. We include results according to our model (labeled `emt' in the figure), and according to the equations of \citet{2007ApJ...667.1170S} and \citet{2016ApJ...825...71D} (labeled `Sepinsky'). The latter equations, for zero ejection/accretion radii, are of the form of equations~(\ref{eq:av_EOM_tau_nonzero}), with $f_{\dot{M}}=1$, $f_a=\sqrt{1-e^2}$, $f_e=\sqrt{1-e^2}(1-e)$, and $f_\omega=0$. The canonical analytic expectation, $M_\d^2M_\a^2a$ is constant, is represented in the top panel of \F\,\ref{fig:binary_circular} with the red dotted line.

In our example in \F\,\ref{fig:binary_circular}, $q>1$ initially, and the orbit shrinks accordingly. After $\approx 10\,\mathrm{Myr}$, $q$ reaches unity (see the black horizontal dotted line in the bottom panel of \F\,\ref{fig:binary_circular}). Subsequently, the orbit expands. The middle panel illustrates the issue with the equations of motion of \citet{2007ApJ...667.1170S} and \citet{2016ApJ...825...71D}, as mentioned in the Introduction: in the `Sepinsky' model, the eccentricity becomes negative, and subsequently grows to $\gtrsim 0.6$ by 70 Myr. Since $\langle \dot{a} \rangle$ depends on $e$, this also slightly affects the evolution of $a$ after $\approx 50 \, \mathrm{Myr}$. In contrast, the eccentricity in the `emt' model remains zero the entire time, and the semimajor axis evolution is in accordance with conservation of $M_\d^2M_\a^2a$.

Note that, as shown in the bottom panel of \F\,\ref{fig:binary_circular}, $\EA_0=\pi$ at all times in the `emt' model. This reflects the fact that the donor is transferring mass in a circular orbit during all orbital phases (initially, $x\simeq 0.086$). 

We emphasize that $\langle \dot{e} \rangle<0$ for $e=0$ according to the Sepinsky model if $q>1$. If $q<1$, then the Sepinsky model gives $\langle \dot{e} \rangle>0$, but this still gives evolution inconsistent with the canonical equation~(\ref{eq:a_dot_can}) for circular orbits. We show an example in \F\,\ref{fig:binary_circular_low_q}, in which the same system is taken as in \F\,\ref{fig:binary_circular}, except that the initial masses are now $M_\d=1.5\,\msun$ and $M_\a=2\,\msun$, such that initially $q=0.75<1$. In this case, the Sepinsky model predicts a growth of eccentricity, with $e$ increasing to $\sim 0.7$ by 100 Myr. Similarly to the case $q>1$, the semimajor axis evolution according to the Sepinsky model deviates from equation~(\ref{eq:a_dot_can}). The `emt' model yields zero eccentricity at all times, with the semimajor axis increasing in accordance with the canonical expectation.

\subsection{Eccentric orbit}
\label{sect:appl_bin:ecc}
\subsubsection{Zero ejection/accretion radii}
\label{sect:appl_bin:ecc:zero}
Next, we consider an eccentric orbit, still setting $\rAd=\rAa=\ve{0}$. The initial parameters are $M_\d=8\,\msun$, $M_\a=1.4\,\msun$, $\langle\dot{M}_\d \rangle= -10^{-8}\,\msun\,\mathrm{yr^{-1}}$, $a=1\,\au$, $e=0.92$, and $R=10\,\rsun$. This can be representative of a binary in which the primary star underwent a supernova explosion, leaving a neutron star (the accretor). In our scenario, the sudden mass loss and velocity kick produced an eccentric (but still bound) orbit, triggering RLOF of the companion star (the donor). The chosen initial parameters are just sufficient to trigger RLOF near periapsis in our model, i.e., $\EA_0 \simeq 0.13>0$. 

\begin{figure}
  \center
  \includegraphics[width=0.52\textwidth,trim=8mm 0mm 0mm 0mm]{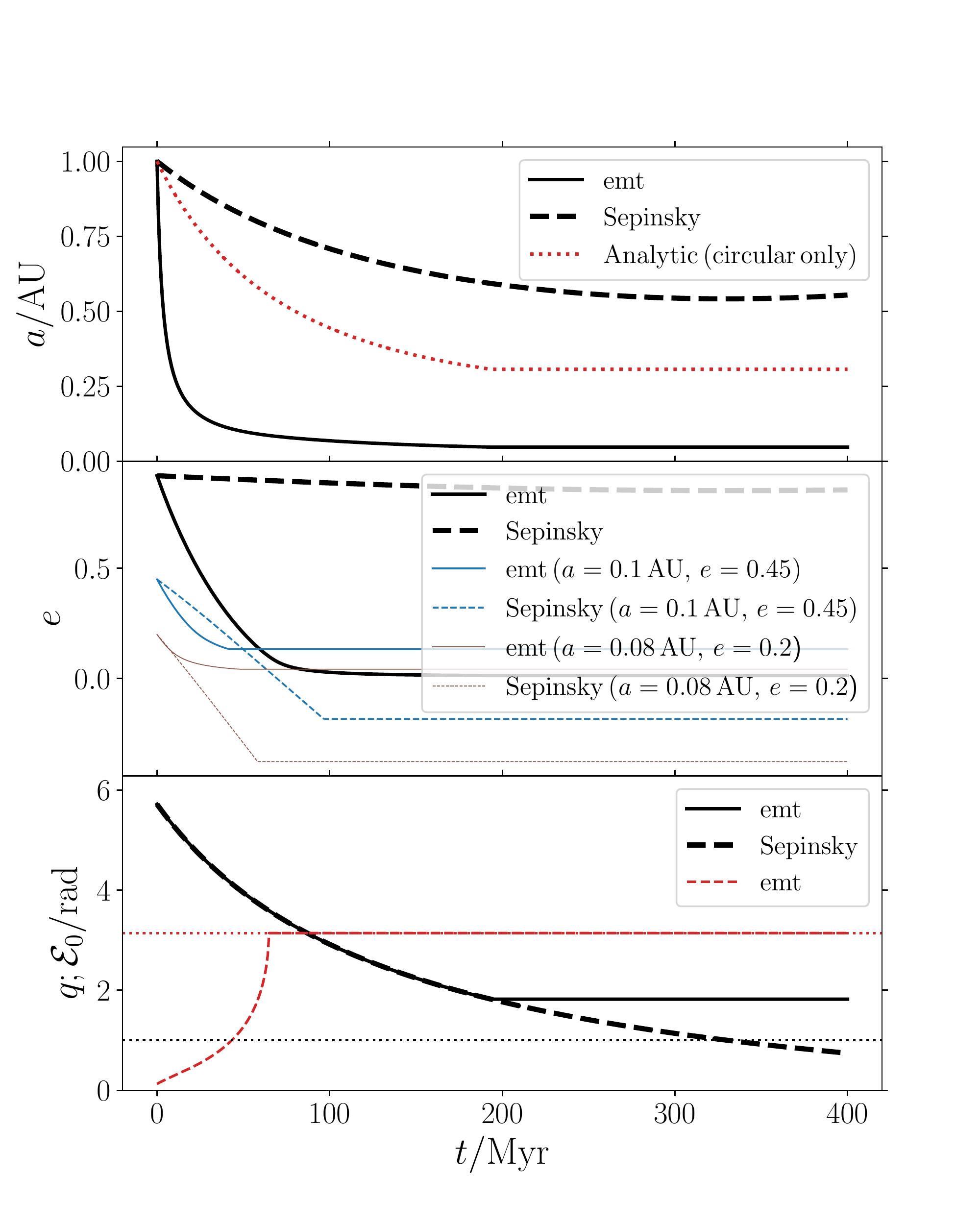}
  \caption{ Similar to \F\,\ref{fig:binary_circular}, now for an eccentric orbit (see \S\,\ref{sect:appl_bin:ecc:zero}). Results are again compared between the `Sepinsky' and `emt' models, and $\rAd=\rAa=\ve{0}$. In the `emt' model, $r_\mathrm{p}=a(1-e)$ reaches $r_\mathrm{p}\leq R$ at $\approx 200\,\mathrm{Myr}$, after which all quantities are taken to be constant. In the middle panel, the blue and brown lines show the eccentricity evolution for two modified cases with lower initial eccentricities, and where $a$ is decreased as well to ensure RLOF occurs during (part of) the orbit. These cases correspond to $a=0.1\,\au$, $e=0.45$ ($a=0.08\,\au$, $e=0.2$) for the blue and (thinner) brown lines, respectively. }
\label{fig:binary_eccentric}
\end{figure}

The orbital evolution is shown in \F\,\ref{fig:binary_eccentric}. According to the `Sepinsky' model, $a$ decreases to $\simeq 0.5\,\au$ after $\approx 400\,\mathrm{Myr}$, whereas the eccentricity decreases slightly. In contrast, according to the `emt' model, the semimajor axis decreases significantly, and the orbit circularizes. Gradually, $\EA_0$ increases, until around 60 Myr, $\EA_0=\pi$, and RLOF occurs during the entire orbit. At $t \approx 200 \, \mathrm{Myr}$, the periapsis distance $r_\mathrm{p}=a(1-e)$ reaches $r_\mathrm{p}\leq R$, in which case we no longer integrate the equations of motion (subsequently, all quantities are taken to be constant). In reality, strong interactions would likely play an important role well before $r_\mathrm{p}\leq R$ is reached. This shows an important qualitative difference between the `Sepinsky' and `emt' models: the `emt' model predicts a strong interaction to occur such as a merger, collision, or strong tidal effects (the latter are not the focus here and so are not included), whereas according to the `Sepinsky' model, no such strong interactions occur. 

In the middle panel of \F\,\ref{fig:binary_eccentric}, we also show with blue and brown lines the eccentricity evolution for two modified cases with smaller initial eccentricities, and where $a$ was decreased as well to ensure RLOF still occurs during (part of) the orbit. These cases correspond to $a=0.1\,\au$, $e=0.45$ ($a=0.08\,\au$, $e=0.2$) for the blue and (thinner) brown lines, respectively. In the `emt' model, the eccentricity and semimajor axis decrease smoothly until $r_\mathrm{p}\leq R$. In the `Sepinsky' model, the decrease in $a$ is smaller, and no collision occurs initially. The eccentricity decreases until $e=0$, after which $e<0$. The integration is halted (constant $e$) when $r_\mathrm{p}\leq R$. 

These examples show that our model deviates significantly from the `Sepinsky' model not only for circular orbits, but for more eccentric orbits as well. The latter can be understood by noting that, even for eccentric orbits, mass transfer in the `emt' model occurs during a finite range of orbital phases, whereas the `Sepinsky' delta function mass transfer rate applies strictly in the limit $e\rightarrow 1$. 

\begin{figure}
  \center
  \includegraphics[width=0.52\textwidth,trim=8mm 0mm 0mm 0mm]{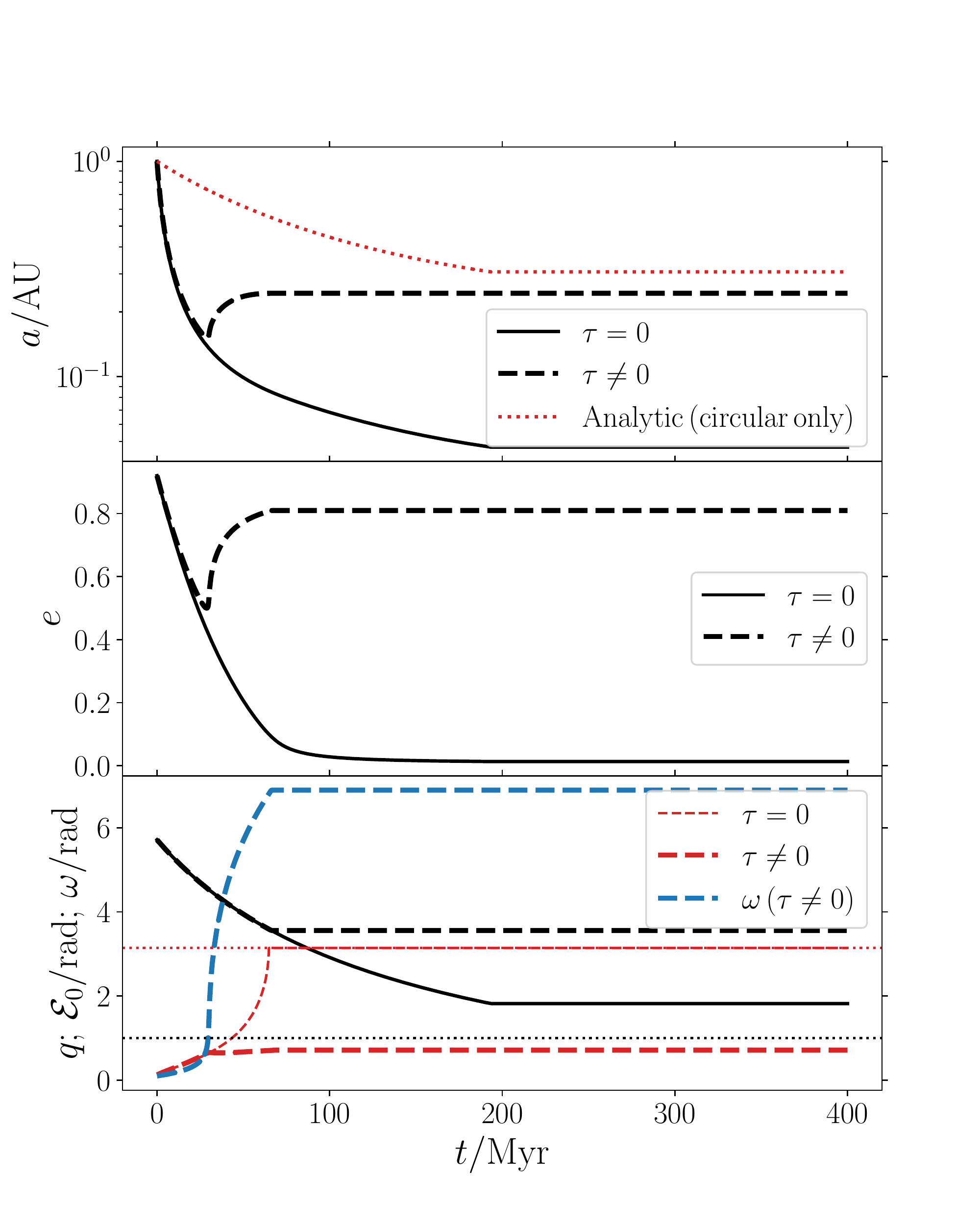}
  \caption{ Similar to \F\,\ref{fig:binary_eccentric}, now comparing results from the `emt' model setting $\tau=0$, and $\tau=10\,\tau_\mathrm{hyd} \simeq 0.26\,\mathrm{d}$ (see \S\,\ref{sect:appl_bin:ecc:zero}). In the bottom panel, the blue dashed line shows the argument of periapsis, $\omega$, as a function of time for the case $\tau\neq0$. }
\label{fig:binary_eccentric_tau}
\end{figure}

We show the possible effect of a nonzero $\tau$ in \F\,\ref{fig:binary_eccentric_tau}, where, for the same parameters as above (with $a=1\,\au$ and $e=0.92$), we compare the cases $\tau=0$, and $\tau\neq0$. The mass-loss delay time $\tau$ is expected to be of the order of the hydrostatic timescale $\tau_\mathrm{hyd}$ (e.g., \citealt{2016MNRAS.455..462V}), and the latter can be estimated by \citep{2012sse..book.....K} $\tau_\mathrm{hyd} \approx \sqrt{R^3/(GM_\d)} \simeq 0.026\,\mathrm{d}$. Here, we set $\tau$ to $\tau=10\,\tau_\mathrm{hyd}$ (setting $\tau=\tau_\mathrm{hyd}$ did not yield noticeable differences between the cases $\tau=0$ and $\tau\neq0$). As shown in \F\,\ref{fig:binary_eccentric_tau}, for a nonzero $\tau$, the orbit starts expanding again after significant shrinkage of the orbit, and the eccentricity increases. At $\approx 70\,\mathrm{Myr}$, $r_\mathrm{p}\leq R$, which is earlier compared to the case $\tau=0$. Also, note that the apsidal line has advanced by nearly $2\pi$ by the time at which $r_\mathrm{p}\leq R$ (see the blue dashed line in the bottom panel of \F\,\ref{fig:binary_eccentric_tau}). 

\subsubsection{Nonzero ejection/accretion radii}
\label{sect:appl_bin:ecc:nonzero}
\begin{figure}
  \center
  \includegraphics[width=0.52\textwidth,trim=8mm 0mm 0mm 0mm]{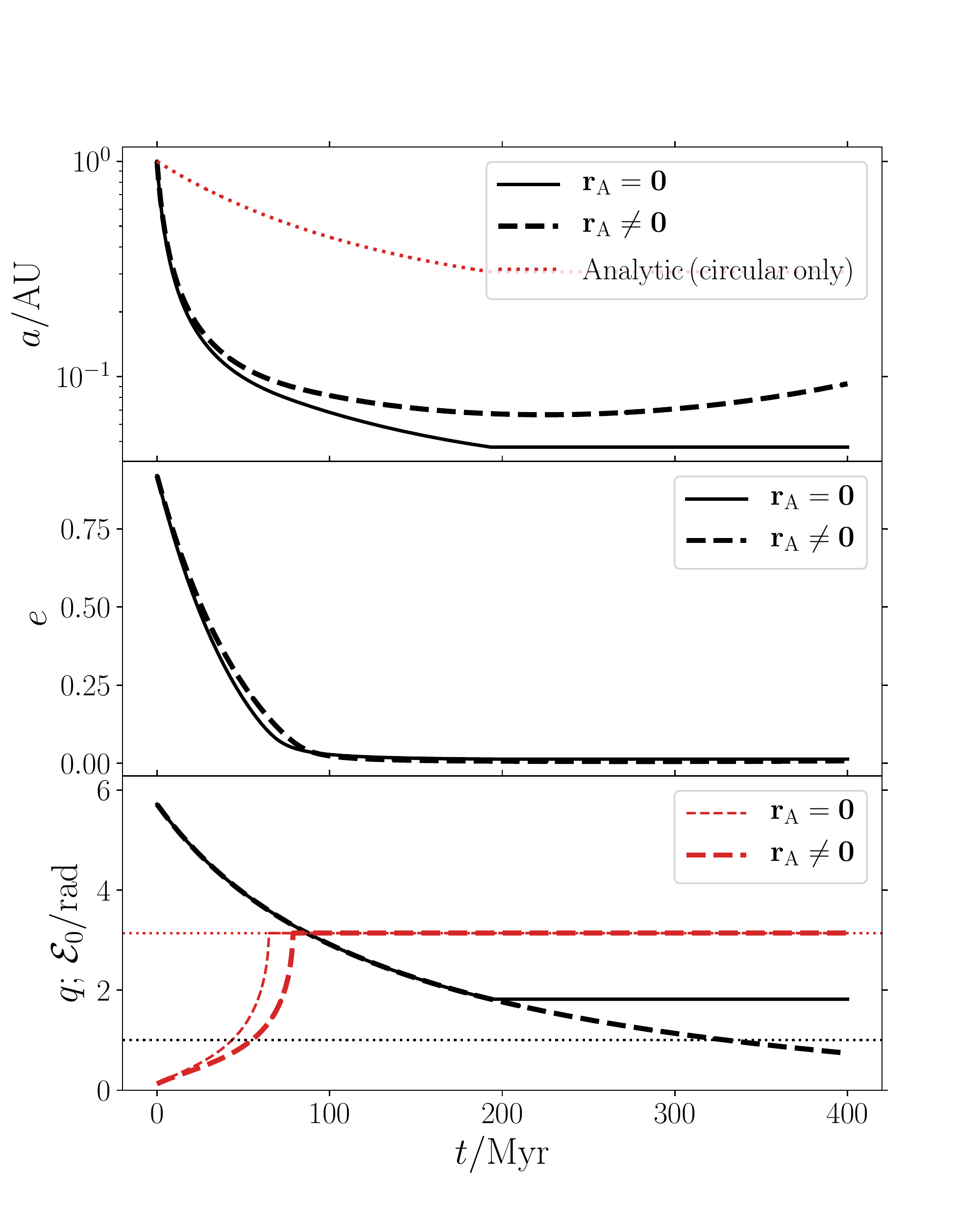}
  \caption{ Similar to \F\,\ref{fig:binary_eccentric}, now comparing results from the `emt' model setting $\rA=\ve{0}$, and $\rA\neq0$ (see \S\,\ref{sect:appl_bin:ecc:nonzero}). }
\label{fig:binary_eccentric_nonzero}
\end{figure}

Next, we focus solely on the `emt' model, and briefly investigate the effect of terms associated with nonzero ejection/accretion radii (i.e., extended bodies). For the eccentric system in \S\,\ref{sect:appl_bin:ecc:zero} with $a=1\,\au$ and $e=0.92$, we compare in \F\,\ref{fig:binary_eccentric_nonzero} the case with $\rA=\ve{0}$ (zero ejection and accretion radii), to $\rA\neq\ve{0}$ (nonzero ejection and accretion radii). In the latter case, we include both a nonzero ejection and accretion radius, where we adopt case (1) from \S\,\ref{sect:model:rA} for the ejection radius (i.e., slow donor spin), and set the accretion radius to $r_\Aa=0.01\,\rsun$ (note that the effect of $r_\Aa$ for our chosen system is very small, unless $r_\Aa\gtrsim1\,\rsun$).

As shown in \F\,\ref{fig:binary_eccentric_nonzero}, the differences between zero and nonzero $\rA$ are not very large. Nevertheless, with $\rA\neq\ve{0}$, the orbit shrinks less, and the collision criterion $r_\mathrm{p}\leq R$ is avoided. Instead, the system evolves to reach mass ratio reversal at $t \approx 330\,\mathrm{Myr}$, around which time the orbit starts expanding. 

\begin{figure}
  \center
  \includegraphics[width=0.52\textwidth,trim=8mm 0mm 0mm 0mm]{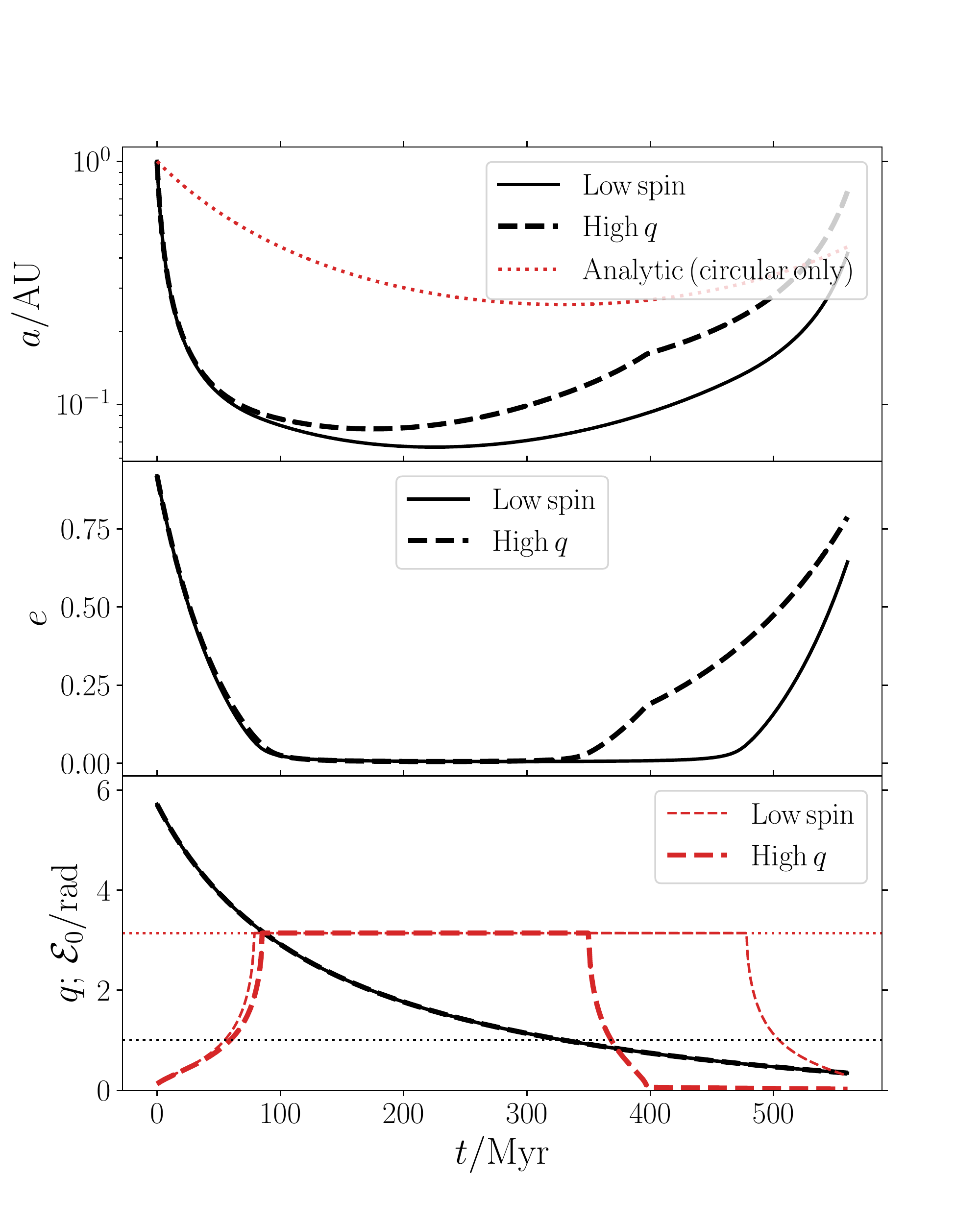}
  \caption{ Similar to \F\,\ref{fig:binary_eccentric}, now applied to the `emt' model with $\rA\neq0$ and comparing the ejection radius models (1) and (2) (see \S\,\ref{sect:appl_bin:ecc:nonzero}). }
\label{fig:binary_eccentric_nonzero_mode}
\end{figure}

Lastly, we compare in \F\,\ref{fig:binary_eccentric_nonzero_mode}, for the same system as above, the cases (1) and (2) for the ejection radius as described in \S\,\ref{sect:model:rA}, i.e., either low donor spin (case 1), or high mass ratio (case 2). In the high-$q$ case, the orbit expands more rapidly after mass ratio reversal. Consequently, mass transfer transitions to partial RLOF around 350 Myr, and the eccentricity starts increasing again. Similar behavior occurs in the low donor spin case, although at later times.

\section{Applications: triple-star systems}
\label{sect:appl_triple}
As demonstrated in \S\,\ref{sect:appl_bin:ecc}, our model already shows more complicated behavior compared to the canonical relation for mass transfer in circular orbits with $\rAd=\rAa=\ve{0}$, i.e., equation~(\ref{eq:a_dot_can}). The latter relation was shown with the red dotted lines in the top panels of the figures. If a tertiary object is included, the realm of possibilities increases even more. Here, we show two illustrative cases, but we emphasize that the parameter space is large; a full investigation is beyond the scope of this paper. Specifically, we consider cases when mass transfer occurs quickly (\S\,\ref{sect:appl_triple:fast}), and slowly (\S\,\ref{sect:appl_triple:slow}). In both cases, we use the code as described in \S\,\ref{sect:model:num}, in which we include the standard secular quadrupole- and octupole-order terms, as well as 1PN terms in the inner orbit. Apart from the 1PN terms, we here focus on mass transfer effects only, and do not include dissipative higher-order PN terms or tides. 

\subsection{Fast mass transfer}
\label{sect:appl_triple:fast}
We choose a triple system with inner binary parameters $M_\d=1\,\msun$, $M_\a=0.1\,\msun$, $\langle \dot{M}_\d \rangle= -10^{-8}\,\msun\,\mathrm{yr^{-1}}$, $a=1\,\au$, $e=0.001$, $\omega=45^\circ$, and $R=1\,\rsun$. The tertiary star has a mass $M_\mathrm{t}=1\,\msun$; the outer orbit has parameters $a_\mathrm{out}=150\,\au$, $e_\mathrm{out}=0.6$, $\omega_\mathrm{out}\simeq 5.7^\circ$, and the mutual inclination with respect to the inner orbit is $i_\mathrm{rel}=85^\circ$. 

\begin{figure}
  \center
  \includegraphics[width=0.52\textwidth,trim=8mm 0mm 0mm 0mm]{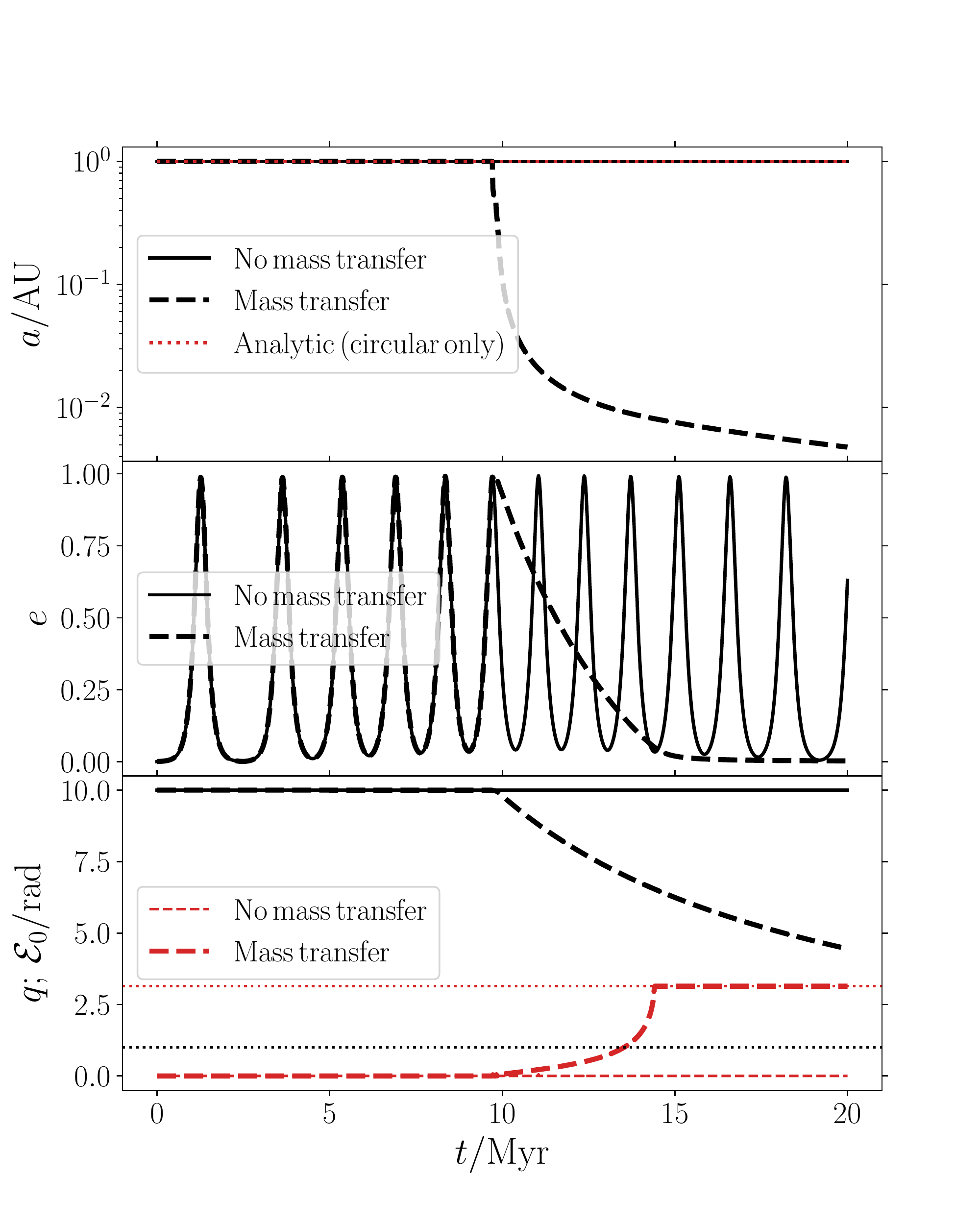}
  \caption{ Evolution of the inner binary semimajor axis (top panel), eccentricity (middle panel), and $q$ and $\EA_0$ (bottom panel), for the triple system discussed in \S\,\ref{sect:appl_triple:fast}. Two cases are shown: with and without mass transfer.  }
\label{fig:triple_fast}
\end{figure}

We carry out two integrations: one with mass transfer included ($\langle \dot{M}_\d \rangle= -10^{-8}\,\msun\,\mathrm{yr^{-1}}$), and one (effectively) without mass transfer ($\langle \dot{M}_\d \rangle= -10^{-30}\,\msun\,\mathrm{yr^{-1}}$). In \F\,\ref{fig:triple_fast}, we compare the two cases. In the absence of mass transfer, the inner orbit eccentricity oscillates with quasi-regular LK oscillations with high amplitude. In the mass transfer case, mass transfer does not occur initially ($\EA_0=0$), but after a few cycles, the eccentricity becomes high enough to trigger partial RLOF ($\EA_0>0$). The semimajor axis quickly decreases (since $q>1$), and the orbit shrinks and circularizes. From the lack of further oscillations after the high-$e$ peak at $\approx 10\,\mathrm{Myr}$, it can be deduced that the inner binary quickly decouples from the tertiary after mass transfer has ensued. 

This type of evolution is analogous to the `fast' type of mergers occurring in black hole triple systems (e.g., \citealt{2018ApJ...864..134R}), or stellar/planetary systems in which tidal friction is important (e.g., \citealt{2015ApJ...799...27P}).

\subsection{Slow mass transfer}
\label{sect:appl_triple:slow}
\begin{figure}
  \center
  \includegraphics[width=0.52\textwidth,trim=8mm 0mm 0mm 0mm]{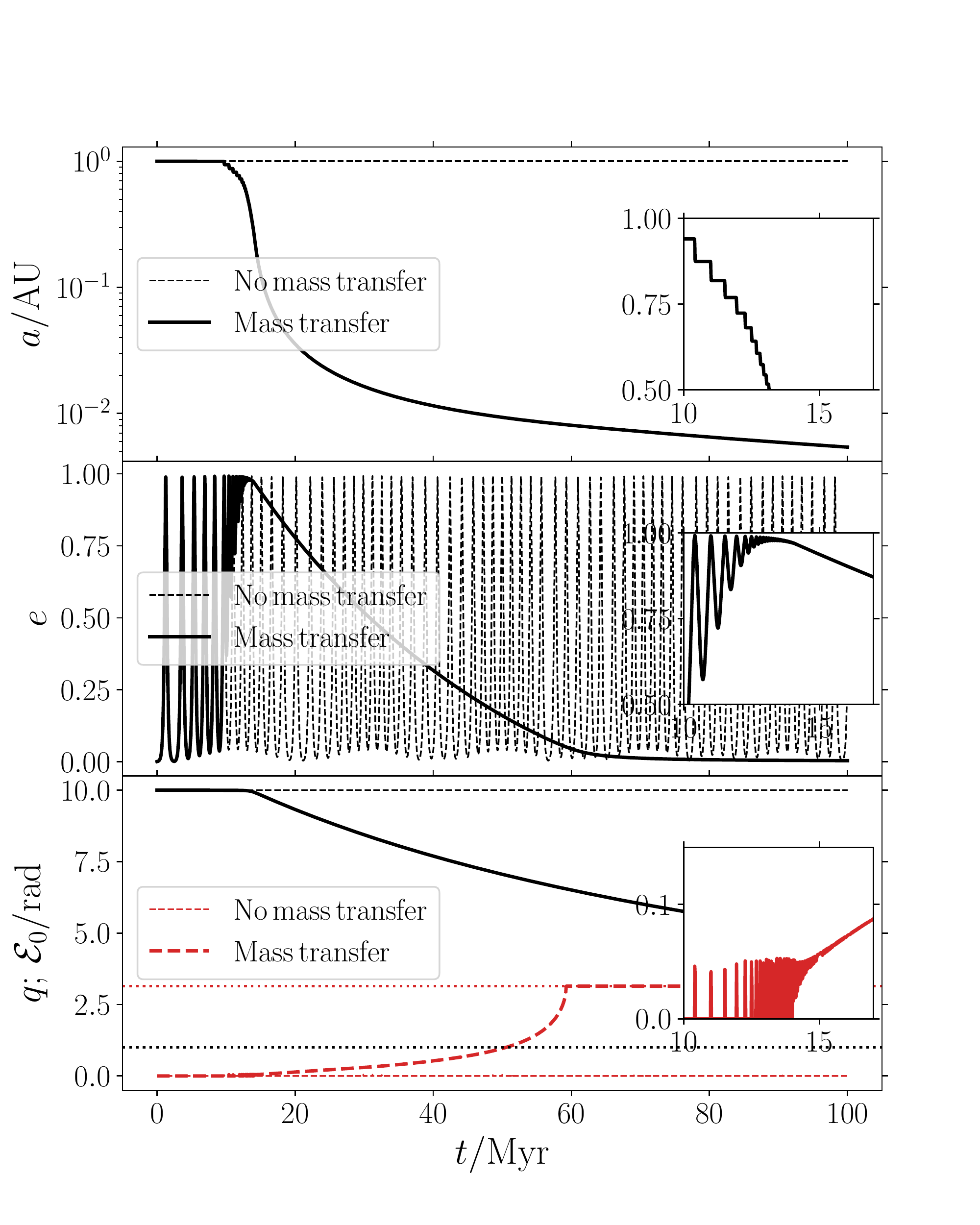}
  \caption{ Similar to \F\,\ref{fig:triple_fast}, now with a higher (absolute) average mass transfer rate, $\langle \dot{M}_\d \rangle= -10^{-9}\,\msun\,\mathrm{yr^{-1}}$. In this case, orbital shrinkage and circularization due to mass transfer occurs more slowly. }
\label{fig:triple_slow}
\end{figure}

In \F\,\ref{fig:triple_slow}, we show the evolution of a very similar system to that in \S\,\ref{sect:appl_triple:fast} -- the only modification is that the average mass transfer rate has been decreased (in absolute value) by a factor of 10, i.e., $\langle \dot{M}_\d \rangle= -10^{-9}\,\msun\,\mathrm{yr^{-1}}$. The resulting evolution is significantly different, however: the inner binary still shrinks and circularizes, but this process takes $\sim 10$ times longer. In particular, the inner binary does not immediately decouple from the tertiary star, as is illustrated with the insets in \F\,\ref{fig:triple_slow}. Initially, around 10 Myr, mass transfer ensues in short bursts at which $\EA_0$ increases to $\approx 0.05\,\mathrm{rad}$ (see the bottom panel of \F\,\ref{fig:triple_slow}). These bursts are associated with stair-wise jumps in the semimajor axis, and gradually locking the eccentricity to a high value. After $\approx 15\, \mathrm{Myr}$, the LK oscillations stop and the orbit circularizes, although partial RLOF continues. After $\approx 60 \, \mathrm{Myr}$, full RLOF occurs ($\EA_0=\pi$ in the bottom panel of \F\,\ref{fig:triple_slow}). 

\section{Discussion}
\label{sect:discussion}
In \S\,\ref{sect:model}, we made a number of simplifying assumptions in order to arrive at an analytically tractable model. These simplifications include the neglect of the force of the mass transfer steam on the stars, conservative mass transfer, imposed ejection and accretion velocities, and specific assumed directions and magnitudes of the ejection/accretion points relative to the donor/accretor. Although most of these assumptions have at least some physical basis, their validity should be evaluated in future work using detailed hydrodynamic simulations. Of course, this endeavor is hampered by the fact that such simulations are typically limited to no more than $\mathcal{O}(10)$ orbits, whereas the secular evolution takes place on much longer timescales. Alternatively, $N$-body simulations tailored for mass transfer to compute the mass transfer stream trajectories and the effect on the orbit could be used to test the long-term evolution, similarly to, e.g., \citet{2010ApJ...724..546S} and \citet{2017ApJ...844...12D}.

Of highest importance is to evaluate the correctness of our assumption of the ejection/accretion velocities, $\ve{w}_\d$ and $\ve{w}_\a$, which leads to the (commonly used) expression equation~(\ref{eq:EOM_simple_zero}) (ignoring back-reaction forces of the matter stream on the stars, and finite sizes of the ejection/accretion radii). More fundamentally, even the physical motivation behind the `canonical' relation for conservative mass transfer in circular binaries, equation~(\ref{eq:a_dot_can}), has been called into question \citep{2008ARep...52..680L}.

\section{Conclusions}
\label{sect:conclusions}
We presented an analytic model to describe the long-term orbital evolution of mass-transferring binaries. Our model applies to conservative mass transfer, in which no net mass is lost from the system. In contrast to previous works, our model is, in principle, applicable for any eccentricity $e$, and it gives qualitatively and quantitively different evolution for both circular and eccentric orbits. We implemented the model in a code that is publicly available (see \S\,\ref{sect:model:num} for the url), and we investigated several cases of Roche Lobe overflow mass transfer in circular and eccentric binaries, and in triples undergoing LK oscillations. Our main conclusions are given below. 

\medskip \noindent 1. The model of \citet{2007ApJ...667.1170S} and \citet{2016ApJ...825...71D} assumed that the mass transfer rate is a delta function of the orbital phase, centered at periapsis. Although physically reasonable in the limit $e\rightarrow1$, this assumption breaks down for smaller eccentricities, when mass transfer is expected to occur for a finite range of orbital phases. In the limit of a circular orbit (in which RLOF occurs during the entire orbit), the mass transfer rate is expected to be constant. As a consequence of the delta function assumption, the equations of motion of the model of \citet{2007ApJ...667.1170S} and \citet{2016ApJ...825...71D} state that the eccentricity time derivative is negative at zero eccentricity (assuming point masses and a donor-to-accretor mass ratio $q>1$), which, when solved (numerically) as a function of time, yields unphysical negative eccentricities (see, e.g., the middle panel of \F\,\ref{fig:binary_circular}). We showed that the delta function model is problematic in the case $q<1$ as well (see \F\,\ref{fig:binary_circular_low_q}). We remedied these issues by assuming that the mass transfer rate is a smooth function of orbital phase, reducing to a delta-like function at high eccentricities, and a flatter function at lower eccentricities. In our model, the eccentricity decays exponentially to zero for small eccentricities (see \S\,\ref{sect:model:prop}). 

\medskip \noindent 2.  We derived explicit expressions for the orbit-averaged equations of motion in our model. We also included the effect of nonzero ejection/accretion radii, making two limiting assumptions related to the location of the ejected material relative to the donor (see \S\,\ref{sect:model:rA}). For zero ejection and accretion radii, we also derived expressions for the orbital evolution when mass transfer occurs with a delay after periapsis. We implemented our expressions into an easy-to-use and freely available \textsc{Python} code to quickly solve the equations of motion numerically, and which can be used as a basis for implementations of our model into binary (and higher-order multiplicity) population synthesis codes. 

\medskip \noindent 3. We applied our model to circular and eccentric binaries. We showed that, in the limit of circular orbits, our model is in agreement with the canonical relation describing conservative mass transfer in circular orbits. In contrast, in our example with a circular system, the model of \citet{2007ApJ...667.1170S} and \citet{2016ApJ...825...71D} yielded an unphysical negative eccentricity, and different semimajor axis evolution compared to the canonical relation (the latter applied to both cases $q>1$, and $q<1$). In examples of eccentric binaries, we found that our model predicts faster orbital shrinkage and circularization compared to the model of \citet{2007ApJ...667.1170S} and \citet{2016ApJ...825...71D}. Also, assuming a delay time between periapsis passage and mass transfer resulted in orbital expansion and eccentricity driving. Furthermore, we found that effects associated with nonzero ejection and accretion radii are somewhat, but not extremely, important. However, we emphasize that we considered only a few examples. A comprehensive parameter space study is beyond the scope of this work.

\medskip \noindent 4. The implementation of our model also allows for inclusion of the (secular) perturbation by a distant orbiting third body. We illustrated the process of mass transfer in triples undergoing LK oscillations in two examples, and showed that mass transfer can act `quickly' (i.e., mass transfer effects are rapidly decoupled from LK oscillations), and `slowly' (in which case many oscillations can occur before the inner binary becomes decoupled). This dichotomy is analogous to triple systems in which dissipative effects are important, such as tidal evolution in systems with planets and/or stars, and gravitational wave emission in systems containing compact objects.

\section*{Acknowledgements}
We thank Scott Tremaine for insightful discussions and comments on the manuscript. We also thank Dimitri Veras for feedback, and the anonymous referee for helpful comments. A.S.H. gratefully acknowledges support from the Institute for Advanced Study, the Peter
Svennilson Membership, and the Martin A. and Helen Chooljian Membership. F.D. acknowledges support from PCTS and Lyman Spitzer Jr fellowships.

\bibliographystyle{yahapj}

\bibliography{literature}

\appendix
\section{A. Explicit expression for the first Lagrange point in the limit of small spin}
\label{app:XL}
If $\mathcal{A}=0$ is set in equation~(\ref{eq:XL}), then the (scaled) first Lagrange point $\XL$ has an analytic solution of $q$, given by
\begin{align}
\XLz(q) = \frac{1}{6} \left(-\sqrt{3} \sqrt{\frac{6 \sqrt{3}
   (q+1)}{\sqrt{A_{-}+A_{+}-2 q+3}}-A_{+}-\frac{q^2}{A_{+}}-4 q+6}+\sqrt{3}
   \sqrt{A_{-}+A_{+}-2 q+3}+3\right).
   \end{align}
Here, 
\begin{align}
A_{\pm} \equiv \sqrt[3]{q \left(q^2\pm6 \sqrt{3} \sqrt{q^2+27}+54\right)}.
\end{align}

\section{B. Explicit expressions for the functions appearing in the orbit-averaged equations of motion}
\label{app:funcs}
Here, we give explicit expressions for the various functions that describe the orbit-averaged equations of motion, and which are referred to in \S\,\ref{sect:model:av}. 
\subsection{Normalization}
The normalization function is obtained by orbit-averaging $\dot{M}_\d$, and reads, for $\tau=0$,
\begin{align}
\label{eq:f_M}
 f_{\dot{M}} (e,x) &=
\nonumber -\frac{1}{96 \pi } \Biggl [ 36 e^4 \EA_0 x^3+3 e^4 x^3 \sin (4 \EA_0)-32 e^3 x^3 \sin (3 \EA_0)+24 e^3 x^2 \sin (3 \EA_0)+288 e^2 \EA_0 x^3-432 e^2
   \EA_0 x^2\\
\nonumber    &\quad+24 e^2 x \left(\left(e^2+6\right) x^2-9 x+3\right) \sin (2 \EA_0)-24 e \left(4 \left(3 e^2+4\right) x^3-9 \left(e^2+4\right)
   x^2+24 x-4\right) \sin (\EA_0)+144 e^2 \EA_0 x\\
   &\quad+96 \EA_0 x^3-288 \EA_0 x^2+288 \EA_0 x-96 \EA_0\Biggl ].
   \end{align}
For $\tau\neq0$, the normalization function reads
\begin{align}
\label{eq:f_M_tau}
\nonumber f_{\dot{M}} (e,x,\EA_\tau) &=-\frac{1}{192 \pi } \Biggl [ 6 e^4 x^3 \sin (2 \EA_0-3 \EA_\tau)+3 e^4 x^3 \sin (4 \EA_0-3 \EA_\tau)+18 e^4 x^3 \sin (2 \EA_0-\EA_\tau)+18 e^4 x^3
   \sin (2 \EA_0+\EA_\tau)\\
\nonumber   &\quad +6 e^4 x^3 \sin (2 \EA_0+3 \EA_\tau)+3 e^4 x^3 \sin (4 \EA_0+3 \EA_\tau)-8 e^3 x^3 \sin (3
   \EA_0-3 \EA_\tau)-72 e^3 x^3 \sin (\EA_0-2 \EA_\tau)-24 e^3 x^3 \sin (3 \EA_0-2 \EA_\tau)\\
\nonumber   &\quad-72 e^3 x^3 \sin   (\EA_0-\EA_\tau)-72 e^3 x^3 \sin (\EA_0+\EA_\tau)-8 e^3 x^3 \sin (3 (\EA_0+\EA_\tau))-72 e^3 x^3 \sin (\EA_0+2
   \EA_\tau)-24 e^3 x^3 \sin (3 \EA_0+2 \EA_\tau)\\
\nonumber   &\quad+72 e^3 x^2 \sin (\EA_0-2 \EA_\tau)+24 e^3 x^2 \sin (3 \EA_0-2 \EA_\tau)+72
   e^3 x^2 \sin (\EA_0+2 \EA_\tau)+24 e^3 x^2 \sin (3 \EA_0+2 \EA_\tau)\\
\nonumber   &\quad+72 e^2 x^3 \sin (2 \EA_0-2 \EA_\tau)+72 e^2 x^3 \sin (2
   \EA_0-\EA_\tau)+72 e^2 x^3 \sin (2 (\EA_0+\EA_\tau))+72 e^2 x^3 \sin (2 \EA_0+\EA_\tau)\\
\nonumber   &\quad-72 e^2 x^2 \sin (2 \EA_0-2
   \EA_\tau)-144 e^2 x^2 \sin (2 \EA_0-\EA_\tau)-72 e^2 x^2 \sin (2 (\EA_0+\EA_\tau))-144 e^2 x^2 \sin (2 \EA_0+\EA_\tau)\\
\nonumber   &\quad+72
   e^2 \EA_0 x \left(\left(e^2+4\right) x^2-8 x+4\right) \cos (\EA_\tau)+72 e^2 x \sin (2 \EA_0-\EA_\tau)+72 e^2 x \sin (2
   \EA_0+\EA_\tau)+288 e^2 \EA_0 x^3-288 e^2 \EA_0 x^2\\
\nonumber   &\quad-96 e (x-1) \left(\left(3 e^2+2\right) x^2-4 x+2\right) \sin (\EA_0)-288 e
   x^3 \sin (\EA_0-\EA_\tau)-288 e x^3 \sin (\EA_0+\EA_\tau)+576 e x^2 \sin (\EA_0-\EA_\tau)\\
   &\quad+576 e x^2 \sin
   (\EA_0+\EA_\tau)-288 e x \sin (\EA_0-\EA_\tau)-288 e x \sin (\EA_0+\EA_\tau)+192 \EA_0 x^3-576 \EA_0 x^2+576
   \EA_0 x-192 \EA_0 \Biggl ].
\end{align}

\subsection{Terms associated with zero ejection/accretion radii}
For $\tau=0$, the functions associated with the terms in $\bf{f}$ proportional to $\dot{\ve{r}}$, which appear regardless of the assumed ejection/accretion locations $\ve{r}_\Ad$ and $\rAa$, are given by
\begin{align}
\nonumber f_a(e,x) &= \frac{1}{96 \pi } \Biggl [ 36 e^4 \EA_0 x^3+3 e^4 x^3 \sin (4 \EA_0)-16 e^3 x^3 \sin (3 \EA_0)+24 e^3 x^2 \sin (3 \EA_0)-144 e^2 \EA_0 x^2+24 e^2 x
   \left(e^2 x^2-3 x+3\right) \sin (2 \EA_0)\\
   &\quad-24 e \left(\left(6 e^2-8\right) x^3+\left(12-9 e^2\right) x^2-4\right) \sin (\EA_0)+144 e^2
   \EA_0 x-96 \EA_0 x^3+288 \EA_0 x^2-288 \EA_0 x+96 \EA_0 \Biggl ]; \\
\nonumber f_e(e,x) &= \frac{1-e^2}{32 \pi } \Biggl [e x \left(12 e^2 \EA_0 x^2+e^2 x^2 \sin (4 \EA_0)+8 \left(\left(e^2+3\right) x^2-6 x+3\right) \sin (2
   \EA_0)-8 e (x-1) x \sin (3 \EA_0)+48 \EA_0 x^2-96 \EA_0 x+48 \EA_0\right)\\
   &\quad-8 (x-1) \left(\left(9 e^2+4\right) x^2-8 x+4\right)
   \sin (\EA_0)\Biggl ].
\end{align}
For $\tau \neq0$, these functions are
\begin{align}
\nonumber f_a(e,x,\EA_\tau) &= \frac{1}{192 \pi } \Biggl [ 6 e^4 x^3 \sin (2 \EA_0-3 \EA_\tau)+3 e^4 x^3 \sin (4 \EA_0-3 \EA_\tau)+18 e^4 x^3 \sin (2 \EA_0-\EA_\tau)+18 e^4 x^3
   \sin (2 \EA_0+\EA_\tau)\\
\nonumber    &\quad +6 e^4 x^3 \sin (2 \EA_0+3 \EA_\tau)+3 e^4 x^3 \sin (4 \EA_0+3 \EA_\tau)+8 e^3 x^3 \sin (3
   \EA_0-3 \EA_\tau)-72 e^3 x^3 \sin (\EA_0-2 \EA_\tau)\\
\nonumber    &\quad -24 e^3 x^3 \sin (3 \EA_0-2 \EA_\tau)+72 e^3 x^3 \sin
   (\EA_0-\EA_\tau)+72 e^3 x^3 \sin (\EA_0+\EA_\tau)+8 e^3 x^3 \sin (3 (\EA_0+\EA_\tau))\\
\nonumber    &\quad-72 e^3 x^3 \sin (\EA_0+2
   \EA_\tau)-24 e^3 x^3 \sin (3 \EA_0+2 \EA_\tau)+72 e^3 x^2 \sin (\EA_0-2 \EA_\tau)+24 e^3 x^2 \sin (3 \EA_0-2 \EA_\tau)\\
\nonumber    &\quad +72
   e^3 x^2 \sin (\EA_0+2 \EA_\tau)+24 e^3 x^2 \sin (3 \EA_0+2 \EA_\tau)-72 e^2 x^3 \sin (2 \EA_0-2 \EA_\tau)+72 e^2 x^3 \sin (2
   \EA_0-\EA_\tau)\\
\nonumber    &\quad-72 e^2 x^3 \sin (2 (\EA_0+\EA_\tau))+72 e^2 x^3 \sin (2 \EA_0+\EA_\tau)+72 e^2 x^2 \sin (2 \EA_0-2
   \EA_\tau)-144 e^2 x^2 \sin (2 \EA_0-\EA_\tau)\\
\nonumber    &\quad+72 e^2 x^2 \sin (2 (\EA_0+\EA_\tau))-144 e^2 x^2 \sin (2 \EA_0+\EA_\tau)+72
   e^2 \EA_0 x \left(\left(e^2+4\right) x^2-8 x+4\right) \cos (\EA_\tau)+72 e^2 x \sin (2 \EA_0-\EA_\tau)\\
\nonumber    &\quad+72 e^2 x \sin (2
   \EA_0+\EA_\tau)-288 e^2 \EA_0 x^3+288 e^2 \EA_0 x^2-96 e (x-1) \left(\left(3 e^2+2\right) x^2-4 x+2\right) \sin (\EA_0)+288 e
   x^3 \sin (\EA_0-\EA_\tau)\\
\nonumber    &\quad+288 e x^3 \sin (\EA_0+\EA_\tau)-576 e x^2 \sin (\EA_0-\EA_\tau)-576 e x^2 \sin
   (\EA_0+\EA_\tau)+288 e x \sin (\EA_0-\EA_\tau)+288 e x \sin (\EA_0+\EA_\tau)\\
   &\quad-192 \EA_0 x^3+576 \EA_0 x^2-576
   \EA_0 x+192 \EA_0 \Biggl ]; \\
\nonumber f_e(e,x,\EA_\tau) &= \frac{1-e^2}{32 \pi } \Biggl [ 12 e \EA_0 x \left(\left(e^2+4\right) x^2-8 x+4\right) \cos (\EA_\tau)+\sin (\EA_0) \Biggl \{ 3 e^3 x^3 \cos
   (\EA_0-3 \EA_\tau)+e^3 x^3 \cos (3 (\EA_0-\EA_\tau))\\
\nonumber   &\quad+6 e^3 x^3 \cos (\EA_0+\EA_\tau)+e^3 x^3 \cos (3
   (\EA_0+\EA_\tau))+3 e^3 x^3 \cos (\EA_0+3 \EA_\tau)-8 e^2 x^3 \cos (2 (\EA_0-\EA_\tau))-8 e^2 x^3 \cos (2
   (\EA_0+\EA_\tau))\\
\nonumber   &\quad+8 e^2 x^2 \cos (2 (\EA_0-\EA_\tau))+8 e^2 x^2 \cos (2 (\EA_0+\EA_\tau))+6 e x \left(\left(e^2+4\right)
   x^2-8 x+4\right) \cos (\EA_0-\EA_\tau)-32 e^2 x^3 \cos (2 \EA_\tau)\\
\nonumber   &\quad+32 e^2 x^2 \cos (2 \EA_\tau)-48 e^2 x^3+48 e^2 x^2+24 e x^3 \cos
   (\EA_0+\EA_\tau)-48 e x^2 \cos (\EA_0+\EA_\tau)+24 e x \cos (\EA_0+\EA_\tau)-32 x^3\\
   &\quad+96 x^2-96 x+32\Biggl \} \Biggl ]; \\
\nonumber f_\omega(e,x,\EA_\tau) &= -\frac{\sqrt{1-e^2}}{32 \pi } x \sin (\EA_\tau) \Biggl [e^2 x^2 \sin (4 \EA_0-2 \EA_\tau)-2 e^2 x^2 \sin (2 (\EA_0-\EA_\tau))-2 e^2 x^2 \sin (2
   (\EA_0+\EA_\tau))+e^2 x^2 \sin (2 (2 \EA_0+\EA_\tau))\\
\nonumber   &\quad-12 e^2 \EA_0 x^2+e^2 x^2 \sin (4 \EA_0)+4 \left(\left(e^2+6\right)
   x^2-12 x+6\right) \sin (2 \EA_0)+24 e x^2 \sin (\EA_0-\EA_\tau)-8 e x^2 \sin (3 \EA_0-\EA_\tau)\\
\nonumber   &\quad+24 e x^2 \sin
   (\EA_0+\EA_\tau)-8 e x^2 \sin (3 \EA_0+\EA_\tau)-24 e x \sin (\EA_0-\EA_\tau)+8 e x \sin (3 \EA_0-\EA_\tau)-24 e x
   \sin (\EA_0+\EA_\tau)\\
   &\quad+8 e x \sin (3 \EA_0+\EA_\tau)-48 \EA_0 x^2+96 \EA_0 x-48 \EA_0\Biggl ].
\end{align}

\subsection{Terms associated with nonzero ejection/accretion radii}
The following functions apply only to the case $\tau=0$. The semimajor axis functions associated with the terms involving nonzero ejection/accretion radii are given by
\begin{align}
\nonumber g_a(e,x) &= \frac{1}{32 \pi } \Biggl [ 4 \EA_0 x \left(e^2 \left(\left(e^2-8\right) x^2+12\right)-8 ((x-3) x+3)\right)+e x \Biggl \{e \biggl [8 \left(x \left(\left(e^2+2\right)
   x-6\right)+3\right) \sin (2 \EA_0)+e x (3 e x \sin (4 \EA_0)\\
   &\quad-16 (x-1) \sin (3 \EA_0))\biggl ]-16 x \left(e^2 (x-3)-4 x+6\right) \sin
   (\EA_0)\Biggl \} +64 \sqrt{1-e^2} \tan ^{-1}\left(\sqrt{\frac{1+e}{1-e}} \tan \left(\frac{\EA_0}{2}\right)\right)\Biggl ]; \\
\nonumber h_a(e,x) &= \frac{1}{4 \pi } \Biggl [ e \sin (\EA_0) \left(-\left(e^2-4\right) x^3-\frac{4}{e \cos (\EA_0)-1}-12 x\right)+x \Bigl \{ e^2 (-x) (e x \sin (3 \EA_0)-3 (x-2)
   \sin (2 \EA_0))\\
   &\quad -2 \EA_0 \left(x \left(\left(e^2+2\right) x-6\right)+6\right)\Biggl \}+\frac{8}{\sqrt{1-e^2}} \tan ^{-1}\left(\sqrt{\frac{1+e}{1-e}} \tan \left(\frac{\EA_0}{2}\right)\right) \Biggl ].
\end{align}

Lastly, the eccentricity functions associated with the terms involving nonzero ejection/accretion radii are given by
\begin{align}
\nonumber g_e(e,x) &= \frac{1-e^2}{48 \pi  e} \Biggl [12 \EA_0 \left(e^2 x \left(x \left(\left(e^2+4\right) x-9\right)+6\right)-2\right)+e^2 x \Biggl \{ 6 \left(x \left(2
   \left(e^2+4\right) x-15\right)+6\right) \sin (2 \EA_0)+e x (3 e x \sin (4 \EA_0)\\
   &\quad+(18-20 x) \sin (3 \EA_0))\Biggl \}-6 e \left(x
   \left(e^2 x (14 x-15)+8 (x-3) x+24\right)-4\right) \sin (\EA_0)+48 \sqrt{1-e^2} \tan ^{-1}\left(\sqrt{\frac{1+e}{1-e}} \tan \left(\frac{\EA_0}{2}\right)\right)\Biggl ]; \\
\nonumber h_e(e,x) &= \frac{1}{48 \pi  e} \Biggl [ 144 \left(1-e^2\right)^{3/2} x \tan ^{-1}\left(\sqrt{\frac{1+e}{1-e}} \tan \left(\frac{\EA_0}{2}\right)\right)+48 \sqrt{1-e^2} \left(3
   \left(e^2-1\right) x+1\right) \tan ^{-1}\left(\sqrt{\frac{1+e}{1-e}} \tan \left(\frac{\EA_0}{2}\right)\right)\\
\nonumber   &\quad-\frac{1-e^2}{e \cos (\EA_0)-1} \Biggl \{
   16 e^4 x^3 \sin (2 \EA_0)+4 e^4 x^3 \sin (4 \EA_0)-26 e^3 x^3 \sin (3 \EA_0)+27 e^3 x^2 \sin (3 \EA_0)+24 e^2 \EA_0
   x^3\\
\nonumber   &\quad+60 e^2 x^3 \sin (2 \EA_0)-36 e^2 \EA_0 x^2-126 e^2 x^2 \sin (2 \EA_0)-12 e \EA_0 \left(e^2 x^2 (2 x-3)-2\right) \cos
   (\EA_0)-3 e \biggl [2 \left(7 e^2+8\right) x^3\\
   &\quad-3 \left(3 e^2+16\right) x^2+48 x-8\biggl ] \sin (\EA_0)+72 e^2 x \sin (2 \EA_0)-24
   \EA_0 \Biggl \} \Biggl ].
\end{align}

\end{document}